\documentclass[prb,aps, 
twocolumn,
showpacs,footinbib,amsmath,amssymb,superscriptaddress,10pt]{revtex4-2}
\usepackage[utf8]{inputenc}
\usepackage{subfiles}
\usepackage{float}
\usepackage{graphicx}
\usepackage{dcolumn}
\usepackage{color}
\usepackage{bm}
\usepackage{natbib,hyperref}
\usepackage{amsmath,amssymb,amsfonts,bbold}
\usepackage[normalem]{ulem}
\usepackage{physics}
\usepackage{listings}
\usepackage[dvipsnames]{xcolor}
\usepackage{color}
\usepackage{braket}
\usepackage{mleftright}
\usepackage{xcolor}
\usepackage{comment}

\newcommand{\of}[1]{\left({#1}\right)}
\newcommand{\ofs}[1]{\left[{#1}\right]}
\newcommand{\ofc}[1]{\left\{{#1}\right\}}

\newcommand{\figpanel}[1]{(#1)}

\newcommand{\citeexp}{\cite{Baumann2015, Natterer2017, Choi2017, Yang2017, Willke2018Probing, Willke2018Hyperfine, Bae2018, Willke2019Tunning,Willke2019Magnetic,Yang2019Tuning, Yang2019Coherent, Seifert2020Single, Seifert2020Longitudinal, Wang2023Universal, VanWeerdenburg2021, Steinbrecher2021, Zhang2021,Kim2022, Kovarik2022, Kot2023, Phark2023Double, Phark2023Electric, Wang2023AnAtomic, Zhang2026_ACS_Nano, Bui2024, Kovarik2024, Czap2025, Zhang2026_ACS_Nano, Huang2025, Greule2026,zhang_influence_2023}}
\newcommand{\citetheo}{\cite{Shakirov2019, Lado2017, Reina-Galvez2019, Ast2024, Ye2024,Reina-Galvez2021,Reina-Galvez2023,Reina-Galvez2025}}

\begin{document}
\title{The effects of an alternating bias on a single orbital spin impurity}
\author{Matyas Nachtigall}
\email{matyas.nachtigall@ehu.eus}
\affiliation{Donostia International Physics Center (DIPC),  20018 Donostia-San Sebasti\'an, Spain}
\author{Jose Reina-G{\'{a}}lvez}
\email{jose.reina-galvez@uni-konstanz.de}
\affiliation{University of Konstanz, D-78457 Konstanz, Germany}
\author{Christoph Wolf}
\email{wolf.christoph@qns.science}
\affiliation{Center for Quantum Nanoscience, Institute for Basic Science (IBS), Seoul 03760, Korea}
\affiliation{Ewha Womans University, Seoul 03760, Korea}
\author{Nicol{\'a}s Lorente}
\email{nicolas.lorente@ehu.eus}
\affiliation{Donostia International Physics Center (DIPC),  20018 Donostia-San Sebasti\'an, Spain}
\affiliation{Centro de F{\'{i}}sica de Materiales
        CFM/MPC (CSIC-UPV/EHU),  20018 Donostia-San Sebasti\'an, Spain}

\begin{abstract}
Electron spin resonance with the scanning tunneling microscope (ESR-STM) enables coherent manipulation and detection of individual atomic spins with atomic-scale spatial resolution, yet the microscopic mechanism by which the radiofrequency (RF) bias drives the spin remains under debate. We study this problem using a Redfield quantum-master-equation framework in which the oscillating bias enters explicitly through the electronic reservoirs rather than through an effective time-dependent tunneling amplitude. In the semiclassical limit, the RF voltage periodically sweeps the system through its static transport landscape, so that the transport regime depends jointly on the DC bias and RF amplitude. When the RF trajectory approaches or crosses a charge transition, the damping-like torque (DLT) associated with spin-polarized charge transfer is strongly enhanced. By contrast, the field-like torque (FLT) induced by the exchange field remains finite deeper in the Coulomb-blockade (CB) regime because it originates from virtual charge fluctuations. The RF trajectory also averages strongly bias-dependent static quantities over a finite voltage range, smoothing features such as the Lamb shift near the charge thresholds. These results establish the RF amplitude as a control parameter not only for the strength of the ESR drive, but also for the transport regime and the ratio between FLT and DLT.

\end{abstract}

\date{\today}

\maketitle

\section{Introduction}

Electron spin resonance implemented in a scanning tunneling microscope (ESR-STM) enables electrical manipulation and detection of individual spins with atomic-scale spatial resolution~\citeexp. The technique combines the spatial selectivity of STM with coherent control of atomic and molecular spin states on surfaces. A localized spin quantum impurity (QI) is placed in the junction between a surface and a spin-polarized tip under an external magnetic field, while a radiofrequency (RF) voltage is applied across the junction, Fig.~\ref{fig:diagram} (a). When the RF frequency approaches the Zeeman splitting or any spin-flip-excitation energy of the impurity, the spin dynamics modify the tunneling current and produce an ESR signal. Despite extensive experimental progress, the microscopic process that couples the RF electric field to the localized spin remains under discussion, and several mechanisms have been proposed~\citetheo. Broader discussions can be found in Ref.~\onlinecite{Delgado2021} and, more recently, in Ref.~\onlinecite{Taherpour2026Theory}.

Some of us previously described an ESR driving mechanism based on a time-dependent exchange field~\cite{Reina-Galvez2025,Zhang2026_ACS_Nano}. Coupling to a spin-polarized electrode generates an effective exchange magnetic field through virtual charge fluctuations~\cite{Braun2004,Konig2005Exchange}. If the coupling changes in time, the exchange field acquires an oscillating component that drives the impurity through a field-like torque (FLT)~\cite{Reina-Galvez2019,Kovarik2024,Reina-Galvez2025}. Spin-polarized tunneling also generates a spin-transfer torque~\cite{Reina-Galvez2025,Shakirov2019}. Within this description, its damping-like component (DLT) originates from the time-dependent nonequilibrium spin accumulation produced by repeated tunneling events. The FLT acts coherently through the effective exchange field, whereas the DLT drives the impurity toward a polarized nonequilibrium state and is intrinsically dissipative~\cite{Kovarik2024,Ruckert2026}. Both contributions have been identified experimentally in a pentacene molecule~\cite{Kovarik2024}, and recent measurements have provided direct evidence for the exchange field and its characteristic dependence on DC bias~\cite{Zhang2026_ACS_Nano,Greule2026}.

Previous theoretical treatments have often represented the RF excitation through an effective modulation of the tunneling strength~\cite{Reina-Galvez2019,Reina-Galvez2021,Reina-Galvez2023,Reina-Galvez2025,Zhang2026_ACS_Nano,zhang_influence_2023,Czap2025,Corina_TMPc_paper_2025,Ruckert2026}. The oscillating bias has also been introduced directly into the electronic reservoirs of Anderson-impurity models, either without spin-polarized electrodes~\cite{Jauho1994} or without the exchange magnetic field~\cite{Shakirov2019,Ast2024}. Ye~\textit{et al.}~\cite{Ye2024} included both ingredients numerically, using a hierarchical equations-of-motion approach with an analytical expression to rationalize the FLT regime.

Here, we extend our QME approach to fully include the non-equilibrium aspects of the transport and the driving process. This way, both DLT and FLT follow from the same transport description.
In the semiclassical limit, the RF voltage can be viewed as a periodic trajectory through the static transport landscape,
$V_{\rm DC}\rightarrow V_{\rm DC}+V_{\rm RF}\cos\omega t$.
The RF amplitude can therefore move the system to a mixed transport regime when the instantaneous bias crosses a charge-transition threshold, beyond just modulating the strength of electron transfer. This connects the applied RF voltage directly to the transport regime explored during the cycle and to the relative weight of DLT and FLT.

Section~\ref{sec:theory} extends the Redfield QME by including the RF bias explicitly in the reservoirs. Section~\ref{sec:driving} discusses the resulting spin-driving mechanism, while Sec.~\ref{sec:transprot} classifies the transport regimes in terms of the DC and RF voltages. Section~\ref{sec:results} applies the current treatment to the DLT- and FLT-driven spin dynamics of a spin-1/2 impurity.

\section{Theory}
\label{sec:theory}

\begin{figure*}
    \centering
    \includegraphics[width=0.835\linewidth]{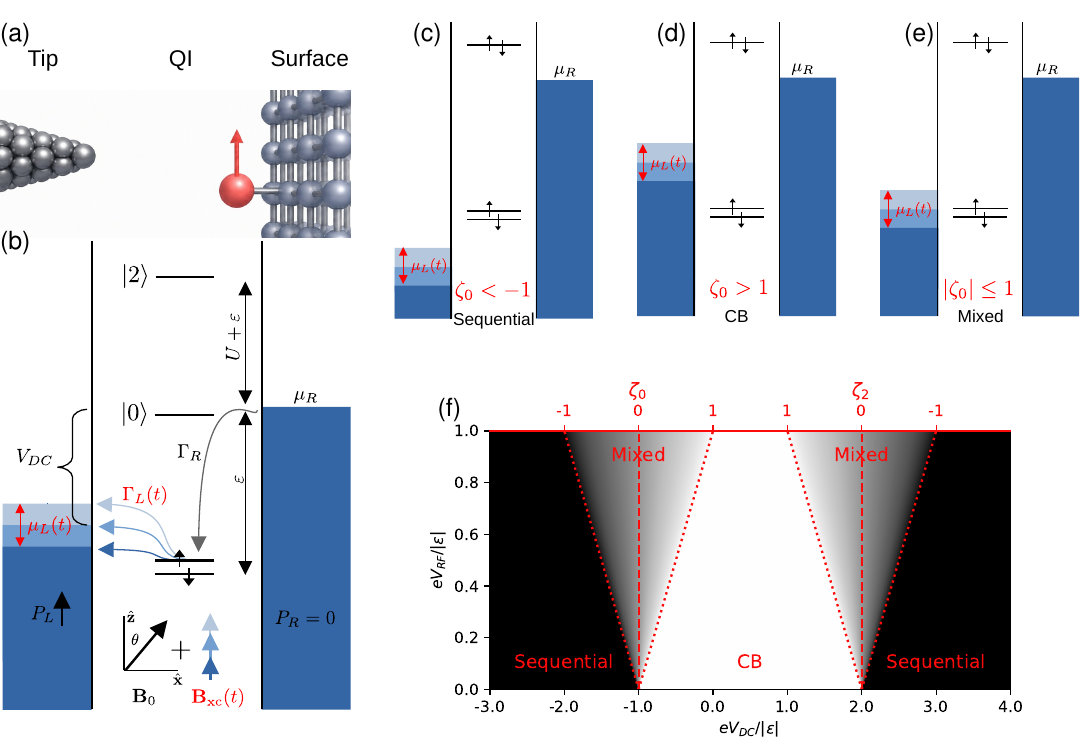}
    \caption{\figpanel{a} Schematic illustration of an ESR-STM experiment with a spin-polarized tip approaching a QI adsorbed on a nonmagnetic surface.
    \figpanel{b} Schematic representation of the theory for a spin-1/2 impurity.
    The impurity is represented by the many-body levels $\ket{0}$, $\ket{2}$, $\ket{\uparrow}$, and $\ket{\downarrow}$, coupled to an unpolarized surface ($P_R=0$) and to a spin-polarized tip pointing along the $z$ direction and having magnitude $P_L$.
    The reservoir Fermi levels are $\mu_R$ and $\mu_L\of{t}$, and the time-dependent bias is $eV(t)=eV_{\rm DC}+eV_{\rm RF}\cos\omega t=\mu_L\of{t}-\mu_R$.
    The external magnetic field, $\mathbf{B}_{0}$, lies in the $xz$ plane and forms an angle $\theta$ with the $z$ axis.
    $\mathbf{B}_{0}$ is supplemented by an effective exchange magnetic field $\mathbf{B}_{xc}\of{t}$ generated by coupling to the spin-polarized electrode.
    Oscillations of the left-electrode Fermi level render the transition rates involving the tip time dependent, producing both a time-dependent exchange field $B_{xc}\of{t}$ and a time-dependent spin-accumulation term. These two contributions generate the FLT and DLT driving mechanisms, respectively.
    \figpanel{c}--\figpanel{e} Schematic illustration of the transport regimes around the negative-bias charge transition.
    In the sequential regime \figpanel{c}, $\zeta < -1$, the Fermi levels satisfy one of the following statements $U+\varepsilon < eV_{\rm DC} + eV_{\rm RF}\cos\omega{t}$ or $ eV_{\rm DC} + eV_{\rm RF}\cos\omega{t}<\varepsilon$ at all times. 
    In the CB regime \figpanel{d}, $\zeta > 1$, the instantaneous bias remains within the charging window, $\varepsilon<eV(t)<\varepsilon+U$, throughout the cycle.
    In the mixed regime \figpanel{e}, $\abs{\zeta}\leq1$, the RF trajectory crosses the charge-transition threshold at $eV(t)=\varepsilon$ or $eV(t)=\varepsilon+U$ during each cycle and alternates between the sequential and CB regimes.
    \figpanel{f} Map of the different transport regimes in the $V_{\rm DC}$--$V_{\rm RF}$ plane. The conditions $\zeta_{0,2}=\pm1$ define the boundaries between the regimes.
    }
    \label{fig:diagram}
\end{figure*}

We consider a QI coupled to two electronic reservoirs, representing the tip and the surface, through a tunneling Hamiltonian, as shown in Fig.~\ref{fig:diagram}\figpanel{a}.
The total Hamiltonian is
\begin{align}\begin{split}
    \label{eq:hamiltonian}
    H\of{t} &= \sum_\alpha H_\alpha\of{t} + H_{QI} + H_T.
\end{split}\end{align}

The first term describes the non-interacting electronic reservoirs,
\begin{align}\begin{split}
    H_\alpha\of{t} =
    \sum_{ k \sigma}
    \left[
    \xi_{\alpha k \sigma}
    -eV^{\rm RF}_{\alpha}\cos\omega t
    \right]
    c^\dag_{\alpha k \sigma}c_{\alpha k \sigma}.
    \label{eq:bath}
\end{split}\end{align}
Here, $\alpha=L,R$ labels the left (tip) and right (surface) electrodes, respectively, while $k$ and $\sigma$ denote orbital momentum and spin. The static electrochemical potential of electrode $\alpha$ is $\mu_\alpha^0$, with occupation
\begin{align}
f_\alpha^+\of{\epsilon}
=
\frac{1}{e^{(\epsilon-\mu_\alpha^0)/(k_B T)}+1},
\qquad
f_\alpha^-\of{\epsilon}=1-f_\alpha^+\of{\epsilon}.
\end{align}
We take the surface Fermi level as the reference energy, $\mu_R^0=0$, such that $\mu_L^0=eV_{\rm DC}$. The RF voltage is applied only to the tip, $V_L^{\rm RF}=V_{\rm RF}$ and $V_R^{\rm RF}=0$. In Eq.~\eqref{eq:bath}, the RF voltage produces a rigid time-dependent shift of the single-particle energies in the corresponding electrode and therefore leaves its particle number unchanged.

Within the wide-band approximation, the spin-dependent density of states $D_{\alpha\sigma}$ is assumed constant over the relevant energy window and zero outside a cutoff $\pm E_{\rm cut}$. We define the total density of states as $D_\alpha=D_{\alpha\uparrow}+D_{\alpha\downarrow}$ and the tip spin polarization as $P_\alpha =    \frac{D_{\alpha\uparrow}-D_{\alpha\downarrow}}{D_\alpha}.$ As in ESR-STM experiments, we assume that only the tip electrode is polarized, $P_L\neq 0$, whereas the substrate is unpolarized, $P_R=0$.

$H_{QI}$ describes a quantum impurity model introduced in Refs.~\onlinecite{Reina-Galvez2021,Reina-Galvez2023,Reina-Galvez2025},
\begin{align}\begin{split}
    H_{QI}
    =&
    \sum_{\sigma=\uparrow,\downarrow}
    \varepsilon\hat{d}^{\dag}_\sigma\hat{d}_\sigma
    +U\hat{n}_{\uparrow}\hat{n}_{\downarrow}\\
    &+g\mu_B\mathbf{B}_{0}\cdot
    \of{\hat{\mathbf{S}}+\hat{\mathbf{s}}}
    +\hat{\mathbf{S}}\cdot J\cdot\hat{\mathbf{s}}.
\end{split}\end{align}
$\hat{d}_\sigma$ annihilates an electron with spin $\sigma$, and $\hat{n}_\sigma=\hat{d}^{\dag}_\sigma\hat{d}_\sigma$ is the corresponding number operator.
We use dimensionless spin operators throughout. The transport-orbital spin is therefore defined as $\hat{\mathbf{s}} =  \sum_{\sigma\sigma'}  \hat{d}^{\dag}_\sigma\boldsymbol{\sigma}_{\sigma\sigma'}  \hat{d}_{\sigma'}/2,$ where $\boldsymbol{\sigma}$ denotes the vector of Pauli matrices. The operator $\hat{\mathbf{S}}$ describes additional localized spin degrees of freedom coupled to the transport orbital. $\varepsilon$ and $U$ are the onsite energy and Coulomb repulsion, respectively. When the cutoff energy, $E_{cut}$, is much larger than all other QI energy scales, i.e. $E_{cut}\gg |\varepsilon|,\ |\varepsilon+U|$, the rates converge to a cutoff-independent value~\cite{Braun2004,Piotr_Martinek_2021_JMMM,Piotr_Martinek_Hanle_effect_2023,Busz_PRB_2025}. This condition is satisfied in all simulations presented below. $J$ is the exchange interaction tensor and $\mathbf{B}_{0}$ is the external magnetic field, forming an angle $\theta$ with the tip-polarization direction.
We choose the coordinate system such that the tip spin polarization points along $\hat{\mathbf{z}}$ and the external magnetic field lies in the $xz$ plane.
The Hamiltonian can be generalized further by including a dipolar field generated by the spin-polarized electrode and magnetic anisotropy induced by the surface environment.

We write the QI Hamiltonian in its many-body eigenbasis,
\begin{align}\begin{split}
    H_{QI} &= \sum_j E_j\ket{j}\bra{j}.
\end{split}\end{align}
The many-body eigenstates $\ket{j}$ follow from diagonalizing $H_{QI}$ in the basis $\ket{0,m}$, $\ket{\sigma,m}$, and $\ket{2,m}$, where $\sigma=\uparrow,\downarrow$ denotes the transport-spin projection along the tip-polarization axis and $m$ denotes the projection of the additional localized spin $\mathbf{S}$. These labels refer to basis states rather than, in general, to eigenstates of the interacting system. The singly occupied eigenstates are generally superpositions of different spin configurations and may instead be labeled according to their energies, for example as ground and excited states~\cite{Reina-Galvez2025}. In the empty and doubly occupied sectors, the transport spin vanishes, $\mathbf{s}=0$, while the localized spin $\mathbf{S}$ remains present.

The tunneling Hamiltonian is
\begin{align}\begin{split}    
    H_T =
    \sum_{\alpha k\sigma lj}
    \left[
    t_\alpha^0
    \lambda_{lj\sigma}
    c^\dag_{\alpha k\sigma}
    \ket{l}\bra{j}
    +\text{h.c.}
    \right],
\end{split}\end{align}
where
\begin{align}
    \lambda_{lj\sigma}
    =
    \langle l|
    \sum_m
    \left(
    |0,m\rangle\langle\sigma,m|
    +
    |\bar{\sigma},m\rangle\langle2,m|
    \right)
    |j\rangle
\end{align}
describes the change in the QI many-body configuration associated with annihilating a transport electron. Its dependence on $\theta$ enters through the many-body eigenstates.
$t^0_\alpha$ is the bare tunneling amplitude between the QI and electrode $\alpha$.

The time-dependent RF contribution can be shifted from the reservoir Hamiltonian to the phase of the tunneling amplitude through the unitary transformation
\begin{align}
    \hat{U}\of{t}
    =
    \exp\left[
    \frac{i}{\hbar}
    \sum_\alpha
    \int_0^t dt'\,
    eV_\alpha^{\rm RF}\cos\of{\omega t'}\hat{N}_\alpha
    \right],
\end{align}
with $\hat{N}_\alpha=\sum_{k\sigma}c^\dag_{\alpha k\sigma}c_{\alpha k\sigma}.$ This transformation removes only the oscillating RF contribution. The DC bias remains encoded in the static chemical potentials $\mu_\alpha^0$ that determine the reservoir occupations.
Following Ref.~\onlinecite{Gonzalez2020}, the transformed reservoir and tunneling Hamiltonians become
\begin{align}\begin{split}
    H'_\alpha
    =&
    \sum_{k\sigma}
    \xi_{\alpha k\sigma}
    c^\dag_{\alpha k\sigma}c_{\alpha k\sigma},
    \\
    H'_T\of{t}
    =&
    \sum_{\alpha k\sigma lj}
    t^0_\alpha
    \lambda_{lj\sigma}
    c^\dag_{\alpha k\sigma}
    \ket{l}\bra{j}
    \sum_n
    J_n\of{\frac{eV^{\rm RF}_\alpha}{\hbar\omega}}
    e^{in\omega t}\\
    &+\text{h.c.}
\end{split}\end{align}
The RF voltage therefore generates photon-assisted tunneling sidebands separated by $\hbar\omega$, with amplitudes weighted by the Bessel functions $J_n(eV_\alpha^{\rm RF}/\hbar\omega)$. Since $V_R^{\rm RF}=0$ for the surface electrode, only its $n=0$ contribution survives.

We describe the QI through a reduced density matrix obtained by tracing over the reservoir degrees of freedom~\cite{Hewson1993,Bhandari2021},
\begin{align}\begin{split}    
    \hat{\rho}\of{t}
    =
    \Tr_R\left[\hat{\rho}_T\of{t}\right]
    =
    \mleft(
    \begin{array}{ccc}
    \rho_{11} & 0 & 0 \\
    0 & \rho_{00} & 0 \\
    0 & 0 & \rho_{22}
    \end{array}
    \mright),
\end{split}\end{align}
where $\hat{\rho}_T$ denotes the density matrix of the complete system.
Its matrix elements in the QI eigenbasis are $\rho_{lj}=\langle l|\hat{\rho}|j\rangle$.
The submatrices $\rho_{pp}$ describe the magnetic states within occupation sector $p$.
If $N$ is the Hilbert-space dimension of the localized spin $\mathbf{S}$, the $p=0$ and $p=2$ blocks have dimension $N\times N$, while the singly occupied block $\rho_{11}$ has dimension $2N\times2N$ because it also contains the transport-spin degree of freedom. In the spin-$1/2$ case, $\rho_{11}$ reduces to the $2\times2$ block containing $\rho_{gg}$, $\rho_{ee}$, and the coherences $\rho_{ge}$ and $\rho_{eg}$. For normal particle-number-conserving reservoirs, coherences between QI states belonging to different charge sectors do not contribute to the reduced dynamics, giving the block-diagonal form above. 

Following Ref.~\onlinecite{Schoeller1994,Konig1996Resonant,Konig1996Zero,Splettstoesser2006,Esposito2009,Cavaliere2009,Reina-Galvez2021,Reina-Galvez2023,Bhandari2021}, we expand the periodic reduced density matrix as
\begin{align}
    \rho_{ij}\of{t}
    =
    \sum_n\rho_{ij;n}e^{in\omega t}.
\end{align}
The resulting Redfield QME is evaluated to lowest order in the tunnel coupling, corresponding to second order in the tunneling Hamiltonian $H_T$ and first order in the hybridization strength.
The equation of motion is (see Appendix~\ref{sec:app-rates})
\begin{align}\begin{split}
    (\Delta_{lj}-n\hbar\omega)\rho_{lj;n}
    &=
    i\hspace{-0.05cm}\sum_{vu;n'}\hspace{-0.05cm}
    \left[
    \left(\Gamma_{vl,ju;n'}    +
    \Gamma^*_{uj,lv;-n'}\right)\hspace{-0.05cm}\rho_{vu;n-n'}\right.\\
    &\left.- \Gamma^{*}_{lv,vu;-n'}\rho_{uj;n-n'}
    -\Gamma_{jv,vu;n'}\rho_{lu;n-n'}
    \right]
\end{split}\end{align}
where $\Delta_{lj}=E_l-E_j$. The transition rates are
\begin{align}\begin{split}
\label{eq:rate-four}
    \Gamma_{vl,ju,\alpha;n}^+
    =&    \sum_{\sigma}
    \gamma^0_{\alpha\sigma}
    \lambda_{vl\sigma}\lambda^*_{uj\sigma}
    \sum_{n'}
    I_{\alpha}^{+}\of{\Delta_{ju}-n'\hbar\omega}\\
    &\times J_{n'}\of{\frac{eV_{\alpha}^{\rm RF}}{\hbar\omega}}
    J_{n'-n}\of{\frac{eV_{\alpha}^{\rm RF}}{\hbar\omega}},
    \\
    \Gamma_{vl,ju,\alpha;n}^-
    =&
    \sum_{\sigma}
    \gamma^0_{\alpha\sigma}
    \lambda^*_{lv\sigma}\lambda_{ju\sigma}
    \sum_{n'}
    I_{\alpha}^{-}\of{\Delta_{ju}+n'\hbar\omega}\\
    &\times J_{n'+n}\of{\frac{eV_{\alpha}^{\rm RF}}{\hbar\omega}}
    J_{n'}\of{\frac{eV_{\alpha}^{\rm RF}}{\hbar\omega}},
    \\
    \Gamma_{vl,ju;n}
    =&
    \sum_\alpha
    \left(
    \Gamma_{vl,ju,\alpha;n}^+
    +
    \Gamma_{vl,ju,\alpha;n}^-
    \right).
\end{split}\end{align}
The terms $\Delta_{ju}\pm m\hbar\omega$ represent transitions through Floquet sidebands displaced from the bare QI transition energy by integer multiples of $\hbar\omega$.
The coupling to electrode $\alpha$ is 
    $\gamma^0_{\alpha\sigma}=    2\pi\abs{t^0_\alpha}^2D_{\alpha\sigma}=    \pi\abs{t^0_\alpha}^2D_\alpha    \of{1+2\sigma P_\alpha}$, where $\sigma=+1/2$ and $-1/2$ correspond to spin up and down, respectively.

The reservoir occupations enter through
\begin{align}\begin{split}
\label{eq:fermi-integral}
    I_{\alpha}^{\pm}\of{\Delta_{ij}}
    =&
    \frac{i}{2\pi}
    \int_{-E_{\rm cut}}^{E_{\rm cut}}
    \frac{f^{\pm}_\alpha\of{\epsilon}}
    {\pm\epsilon-\Delta_{ij}
    +\frac{i\hbar}{\tau_C}}
    d\epsilon.
\end{split}\end{align}
Here, $f_\alpha^+$ and $f_\alpha^-$ describe occupied and empty reservoir states, respectively. The $+$ contribution therefore describes processes in which an occupied reservoir state supplies an electron to the QI, whereas the $-$ term requires an empty reservoir state and corresponds to an electron tunneling from the QI into the electrode.

$\hbar\tau_C^{-1}$ represents a phenomenological broadening of the QI levels that vanishes in the strict lowest-order limit.
We retain a finite value to avoid singularities in $I_{\alpha}^{\pm}\of{E}$.
This quantity may be interpreted as an imaginary contribution to the QI self-energy arising from hybridization with the electrodes. It produces finite spectral tails away from a charge transition but does not constitute an explicit treatment of cotunneling. For a time-independent Anderson impurity coupled to a single electrode, the corresponding hybridization scale is $ \hbar\tau_C^{-1}    =\pi\abs{t^0}^2D_\alpha$~\cite{Yosida1998}.

We define the current flowing from the left electrode into the QI as
\begin{align}    
    I_L\of{t}
    =
    -e\frac{d}{dt}\langle\hat{N}_L\rangle
    =
    \sum_n I_n e^{in\omega t}.
\end{align}
Its Fourier components are
\begin{align}\begin{split}
\label{eq:current}
    I_n
    =
    \frac{e}{\hbar}
    \sum_{lju;m}
    \left[
    \rho_{lu;-n-m}
    \Gamma^C_{lj,ju,L;m}
    +
    \rho^*_{lu;n-m}
    \Gamma^{C*}_{lj,ju,L;m}
    \right].
\end{split}\end{align}
The current follows from the transition rates by assigning opposite signs to processes that transfer an electron into or out of electrode $\alpha$,   $ \Gamma_{vl,ju,\alpha;n}^C=    \Gamma_{vl,ju,\alpha;n}^+    -    \Gamma_{vl,ju,\alpha;n}^-.$ The sign difference reflects the opposite changes in the electrode particle number produced by the two processes.

\subsection{Dynamics of a Spin-1/2 System}
\label{sec:driving}

For the remainder of this work, we focus on a two-level spin system.
We choose the energy of the positively charged, unoccupied state $\ket{0}$ as the reference. The two neutral, singly occupied eigenstates, $\ket{g}$ and $\ket{e}$, have energies $E_g=\varepsilon-\Delta_Z/2$ and $E_e=\varepsilon+\Delta_Z/2$, respectively, where $\Delta_Z=g\mu_B B_0$ is the Zeeman splitting. The negatively charged, doubly occupied state $\ket{2}$ has energy $E_2=2\varepsilon+U$, where $U$ is the Coulomb repulsion.
The reduced density matrix is
\begin{align}\begin{split}
    \hat{\rho}\of{t} =
    \mleft(
    \begin{array}{cccc}
    \rho_{gg} & \rho_{ge} & 0 & 0 \\
    \rho_{ge}^* & \rho_{ee} & 0 & 0 \\
    0 & 0 & \rho_{00} & 0 \\
    0 & 0 & 0 & \rho_{22}
    \end{array}
    \mright).
\end{split}\end{align}

For a spin-1/2 system, it is convenient to describe the singly occupied sector through the dimensionless spin expectation value in a frame whose $z'$ axis is aligned with the external magnetic field $\mathbf{B}_0$. With the phase convention chosen for the eigenstates $\ket{g}$ and $\ket{e}$, the spin components in this frame are
\begin{align}
    \left\langle\hat{\mathbf{s}}'\right\rangle
    &=
    \mleft(
    \begin{array}{c}
    \left\langle{s}'_x\right\rangle\\
    \left\langle{s}'_y\right\rangle\\
    \left\langle{s}'_z\right\rangle
    \end{array}
    \mright)
    =
    \frac{1}{2}
    \mleft(
    \begin{array}{c}
    -\of{\rho_{ge}+\rho_{eg}}\\
    i\of{\rho_{ge}-\rho_{eg}}\\
    \rho_{ee}-\rho_{gg}
    \end{array}
    \mright).
\end{align}
Because the impurity can also occupy the empty and doubly occupied charge states, the total probability of the singly occupied sector satisfies $\rho_{gg}+\rho_{ee}\leq1$. The expectation value above is therefore not normalized within the singly occupied subspace. The primed frame is related to the laboratory frame defined previously by a rotation through the angle $\theta$ about the $y$ axis, with the laboratory $z$ axis pointing along the tip polarization and $\mathbf{B}_0$ lying in the $xz$ plane. Transforming the spin expectation value back to the laboratory frame gives
\begin{align}
    \left\langle\hat{\mathbf{s}}\right\rangle
    =
    \mleft(
    \begin{array}{c}
    \left\langle s'_x\right\rangle\cos\theta
    +\left\langle s'_z\right\rangle\sin\theta\\
    \left\langle s'_y\right\rangle\\
    -\left\langle s'_x\right\rangle\sin\theta
    +\left\langle s'_z\right\rangle\cos\theta
    \end{array}
    \mright).
\end{align}

The dynamics of the transport spin can be written as~\cite{Braun2004,Reina-Galvez2025}
\begin{align}\begin{split}
\label{eq:spin-dyn}
    \frac{d\mathbf{s}\of{t}}{dt}
    =
    \frac{M\of{t}}{\hbar}\hat{\mathbf{z}}
    -
    \frac{\mathbf{s}\of{t}}{\tau_{\rm rel}}
    +
    \frac{g\mu_B}{\hbar}
    \left(
    \mathbf{B}_0+B_{\rm xc}\of{t}\hat{\mathbf{z}}
    \right)
    \cross\mathbf{s} \of{t}.
\end{split}\end{align}
The first term represents spin injection through the spin-polarized electrode. The quantity $M\of{t}$ therefore acts as a source of nonequilibrium spin accumulation along the tip-polarization direction and gives rise to the damping-like torque (DLT)~\cite{Kovarik2024,Ruckert2026}. The second term accounts for spin relaxation caused by tunneling between the singly occupied sector, which carries a finite transport spin, and the empty and doubly occupied sectors, where the transport spin vanishes. This process is characterized by the relaxation time $\tau_{\rm rel}$. Finally, coupling to polarized reservoirs produces spin-dependent virtual charge fluctuations and generates an effective exchange magnetic field $B_{\rm xc}$ along the tip-polarization direction~\cite{Konig2004,Braun2004,Reina-Galvez2025,Ruckert2026}. The component of this field transverse to the transport spin gives rise to the field-like torque (FLT).

Within the present lowest-order transport description, these quantities are given by~\cite{Reina-Galvez2025}
\begin{align}\begin{split}
\label{eq:dynamic-driving}
    M\of{t}
    = &\ 2 \sum_\alpha P_\alpha \bigg[
    \Re\ofc{\Gamma^{+0}_{0gg0,\alpha}\of{t}}\rho_{00}\of{t}\\
    &\hspace{1.05cm}- \Re\ofc{\Gamma^{-0}_{2gg2,\alpha}\of{t}}\rho_{22}\of{t} \\
    &\hspace{1.05cm}+ \Re\ofc{\Gamma^{-0}_{g22g,\alpha}\of{t}
    - \Gamma^{+0}_{g00g,\alpha}\of{t}}
    \frac{\rho_{11}\of{t}}{2}
    \bigg],\\
    \frac{1}{\tau_{\rm rel}}
    =&\ \frac{1}{\hbar}
    \sum_{v=0,2}
    \Re\ofc{\Gamma^{0}_{gvvg}\of{t}+\Gamma^{0}_{evve}\of{t}},\\
    B_{\rm xc}\of{t}
    =&\ \frac{2}{g\mu_B\sin\theta}
    \sum_{v=0,2}
    \Im\ofc{\Gamma_{evvg}\of{t}}.
\end{split}\end{align}
Here, $\Gamma_{vlju,\alpha}^{\pm0}\of{t}$ denotes the corresponding lead-resolved rate evaluated at $P_\alpha=0$, while the polarization dependence is written explicitly through the prefactor $P_\alpha$. The off-diagonal rates $\Gamma_{ev,vg}\of{t}$, which couple the ground and excited spin states, are proportional to both the electrode tip spin polarization $P_\alpha$ and $\sin\theta$. The explicit factor of $\sin\theta$ in the definition above therefore cancels this angular dependence, so that $B_{\rm xc}\of{t}$ itself is independent of $\theta$.

\subsection{Driving Mechanism}

Several microscopic driving mechanisms have been proposed~\cite{Lado2017,Shakirov2019,Reina-Galvez2019,Delgado2021,Ast2024,Ye2024}. In the present treatment, the time dependence arises from the periodic modulation of the electrode energies, or equivalently of the electrochemical-potential difference across the junction. The RF bias makes the transition rates periodic and, as a consequence, renders $M(t)$, $\tau_{\rm rel}(t)$, and $B_{\rm xc}(t)$ time dependent. The oscillating components of $M(t)$ and $B_{\rm xc}(t)$ generate the DLT and FLT driving terms, respectively, whereas the static component of the exchange field modifies the spin splitting.

The QI experiences an effective time-dependent magnetic field
$\mathbf{B}_0+B_{\rm xc}(t)\hat{\mathbf{z}}$.
To first order in the exchange field, the resonance condition reads
\begin{align}
    \hbar\omega=\Delta_Z+\delta E,
    \qquad
    \delta E=g\mu_B B_{{\rm xc};0}\cos\theta,
\end{align}
where $\Delta_Z=g\mu_BB_0$ and $B_{{\rm xc};0}$ denotes the static component of the exchange field.

The oscillating exchange field, together with the time-dependent spin-injection term that produces nonequilibrium spin accumulation, provides an electrical mechanism for ESR without requiring an externally applied time-dependent magnetic field. The localized spin can therefore couple to the RF electric bias through electrically generated spin torques, without invoking a mechanically driven displacement of the impurity as in the piezoelectric mechanism proposed in Ref.~\onlinecite{Lado2017}.

In Ref.~\onlinecite{Reina-Galvez2019,Reina-Galvez2023,Reina-Galvez2021,Reina-Galvez2025}, the time dependence was introduced phenomenologically through a modulation of the tunneling amplitude,
\begin{align}
\begin{split}
    \label{eq:A-tdep}
    \Gamma_{vlju,\alpha}^{\pm}(t)
    \propto
    \of{1+A_\alpha\cos\omega t}^2.
\end{split}
\end{align}
The parameter $A_\alpha$ can effectively account for different microscopic driving mechanisms, such as a bias-induced modulation of the tunneling barrier or a shift of the electrode energies. In those treatments, however, $A_\alpha$ is introduced as an effective parameter that does not depend explicitly on the particular many-body transition or on the applied DC bias.

Here, we instead include the RF modulation of the electrode energies explicitly. More generally, Eq.~\eqref{eq:rate-four} describes photon-assisted tunneling through discrete Floquet sidebands separated by $\hbar\omega$, with each contribution weighted by products of Bessel functions. When $\hbar\omega$ is comparable to the relevant electronic energy scales, the discrete Floquet structure must be retained. In the semiclassical limit $eV_{\rm RF}\gg\hbar\omega$, and provided that the relevant electronic quantities vary smoothly on the scale of $\hbar\omega$, the Floquet description approaches an instantaneous-bias picture~\cite{Moskalets2005}.

In Appendix~\ref{sec:rem-bes-fun}, we show that, in this limit, the time dependence of the rates can be obtained through the replacement
$V_{\rm DC}\rightarrow V_{\rm DC}+V_{\rm RF}\cos\omega t$.
The corresponding rates become
\begin{align}
\begin{split}
\label{eq:gamma-tdep}
    \Gamma_{vlju,\alpha}^+(t)
    =&
    \sum_{\sigma}
    \gamma^0_{\alpha\sigma}
    \lambda_{vl\sigma}\lambda^*_{uj\sigma}
    I_{\alpha}^{+}
    \of{\Delta_{ju}-eV^{\rm RF}_\alpha\cos\omega t},
    \\
    \Gamma_{vlju,\alpha}^-(t)
    =&
    \sum_{\sigma}
    \gamma^0_{\alpha\sigma}
    \lambda^*_{lv\sigma}\lambda_{ju\sigma}
    I_{\alpha}^{-}
    \of{\Delta_{ju}+eV^{\rm RF}_\alpha\cos\omega t}.
\end{split}
\end{align}
The static DC bias remains encoded in the reservoir occupations through the static chemical potentials $\mu_\alpha^0$. Equation~\eqref{eq:gamma-tdep} therefore shows that, in this regime, the RF bias periodically samples the static transport response along the voltage trajectory
$V(t)=V_{\rm DC}+V_{\rm RF}\cos\omega t$.

This instantaneous-bias picture also gives a simple interpretation of the driving mechanism. Since the transition rates depend on the instantaneous voltage, their time dependence can be written as
\begin{align}
\begin{split}
\label{eq:rate-derivative}
    \frac{d\Gamma^{\pm}_{vlju,\alpha}}{dt}
    &=
    \frac{d\Gamma^{\pm}_{vlju,\alpha}}{dV}
    \frac{dV}{dt}
    \\
    &=
    -
    \left.
    \frac{d\Gamma^{\pm}_{vlju,\alpha}}{d V}
    \right|_{V=V(t)}
    V^{\rm RF}_{\alpha}\omega\sin\omega t .
\end{split}
\end{align}
Thus, the RF bias modulates a given transition only when the corresponding static rate depends on the applied voltage.

In the additional weak-RF limit, the same result follows by expanding the transition rate around the static operating point,
\begin{align}
\begin{split}
\label{eq:rate-linear}
    \Gamma^{\pm}_{vlju,\alpha}(t)
    \simeq&
    \Gamma^{\pm}_{vlju,\alpha}\of{V_{\rm DC}}
    +
    V^{\rm RF}_{\alpha}
    \left.
    \frac{\partial\Gamma^{\pm}_{vlju,\alpha}}
    {\partial V}
    \right|_{V=V_{\rm DC}}
    \cos\omega t .
\end{split}
\end{align}
The amplitude of the first-harmonic response is therefore controlled by the bias dependence of the corresponding static quantity. For the spin-injection contribution associated with real charge transfer, the modulation is strongest near charge transitions, where the transition rates change rapidly with bias. The exchange field behaves differently because it originates from virtual charge fluctuations and remains bias dependent within the Coulomb-blockade (CB) regime. Its RF modulation can therefore generate a finite FLT even when the RF trajectory does not reach a charge transition.

Within this semiclassical picture, the time-independent ($n=0$) Fourier component of an observable corresponds to its average over one RF cycle. When written as an average over voltage, this becomes a weighted average of the corresponding static quantity over the interval
$V_{\rm DC}-V_{\rm RF}\leq V\leq V_{\rm DC}+V_{\rm RF}$,
because the sinusoidal RF trajectory spends more time near its turning points.
Finite $V_{\rm RF}$ therefore smooths sharp bias dependences close to the QI charge transitions. This averaging does not remove the underlying singular structure of the lowest-order theory, and a finite broadening $\tau_C^{-1}$ remains necessary to regularize the rates. When this broadening is much smaller than the RF energy scale $eV_{\rm RF}$, the resulting observables depend only weakly on its precise value.

Equation~\eqref{eq:gamma-tdep} and the subsequent instantaneous-bias expressions are not used in the numerical calculations presented below, which employ the full Floquet rates of Eq.~\eqref{eq:rate-four}. They are introduced only to provide an intuitive interpretation of the RF-driven dynamics.

Additional microscopic driving mechanisms, such as a bias-dependent tunneling barrier, can also be incorporated into the present formalism. 
Allowing the tunneling amplitude to depend on the electrode energy, $t_\alpha^0\rightarrow t_\alpha(\varepsilon_k)$, or similarly dropping the wide-band approximation, $D_\alpha\rightarrow D_\alpha(\varepsilon_k)$, leads to a more general expression of electron integrals,
\begin{align}
\begin{split}
    \gamma_{L\sigma} & I_{L}^{\pm}
    \of{\Delta_{ij}\mp eV_{\rm RF}\cos\omega t}
    \rightarrow
    \frac{i}{2} \of{1+2\sigma P_L}\\
    &\times \int_{-\infty}^{\infty}
    D_L\of{\epsilon-eV_{\rm RF}\cos\omega t}
    \abs{t_L\of{\epsilon-eV_{\rm RF}\cos\omega t}}^2
    \\
    &\times
    \frac{
    f^{\pm}_L\of{\epsilon-eV_{\rm RF}\cos\omega t}}
    {\pm\epsilon-\Delta_{ij}+\frac{i\hbar}{\tau_C}}
    d\epsilon.
\end{split}
\end{align}
Any additional variance of the rates with electrode energies will result in further contributions to the driving.

\section{Transport Regimes}
\label{sec:transprot}

Charge transport through the QI is characterized by two main regimes.
Sequential tunneling occurs when a charge transition of the QI lies within the transport window defined by the two electrodes (Fig~\ref{fig:diagram}\figpanel{c}).
Electrons can then tunnel through a real intermediate charge state, producing a finite current already at the lowest order in the tunnel coupling.

When no QI charge transition lies within the transport window, the system is instead in the Coulomb-blockade (CB) regime (Fig~\ref{fig:diagram}\figpanel{d}). In the strict CB limit, sequential charge transport is suppressed, and the remaining current arises from cotunneling processes involving virtual intermediate charge states, which enter at the next order in the tunnel coupling. Virtual charge fluctuations, however, are already present in the lowest-order description. They generate the exchange field discussed above even when the sequential current is strongly suppressed.

The Redfield treatment used here is evaluated to the lowest order in hybridization, corresponding to the second order in the bare tunneling amplitudes. It therefore captures sequential transport but does not explicitly include the cotunneling current. The finite level broadening $\hbar\tau_C^{-1}$ extends the spectral tails of the QI transitions into the CB regime and consequently gives rise to a small current in that regime. This contribution should not be interpreted as an explicit description of cotunneling.

Under RF driving, the instantaneous bias can cross a charge-transition threshold during part of the cycle, resulting in a mixed regime (Fig~\ref{fig:diagram}\figpanel{e}). To characterize this situation, we introduce a dimensionless parameter that measures the distance between the DC operating point and the relevant charge transition in units of the RF trajectory,
\begin{align}\begin{split}
\label{eq:zeta-def}
    \zeta = \begin{cases}
        \zeta_0 = \dfrac{eV_{\rm DC}-\varepsilon}{eV_{\rm RF}},
        &\quad \text{tunneling through $\ket{0}$}, \\[10pt]
        \zeta_2 = \dfrac{\varepsilon+U-eV_{\rm DC}}{eV_{\rm RF}},
        &\quad \text{tunneling through $\ket{2}$}.
    \end{cases}
\end{split}\end{align}

The parameter $\zeta$ is the signed detuning of the DC operating point from the corresponding charge-transition threshold, normalized by the RF amplitude. The definitions of $\zeta_0$ and $\zeta_2$ are chosen so that positive $\zeta$ refers to the Coulomb-blockade (CB) side of the transition and negative $\zeta$ refers to the sequential-tunneling side. When $|\zeta|>1$, the RF trajectory never crosses the transition. For $\zeta>1$, the system remains in the CB regime throughout the cycle, whereas for $\zeta<-1$ it remains in the sequential-tunneling regime. When $|\zeta|<1$, the RF trajectory crosses the charge transition twice per cycle and dynamically alternates between the two transport regimes. The fraction of the cycle spent in the sequential-tunneling regime is
\begin{align}
\label{eq:seq-frac}
    f_{\rm seq}=\frac{1}{\pi}\arccos\zeta.
\end{align}
Thus, positive $\zeta$ corresponds to a trajectory that spends most of the cycle in the CB regime, whereas negative $\zeta$ corresponds to one that spends most of the cycle in the sequential-tunneling regime.

The resulting transport map is illustrated in Fig.~\ref{fig:diagram}\figpanel{f}. At $V_{\rm RF}=0$, the two charge-transition thresholds separate the usual static CB and sequential regimes. Increasing $V_{\rm RF}$ opens finite mixed regions around each threshold, bounded by $\abs{\zeta_{0,2}}=1$, whose width grows directly with the RF amplitude. The RF voltage therefore does more than perturb a fixed transport state: it determines which portion of the transport landscape is visited during each cycle. This regime map provides the basis for interpreting the current and spin-dynamical results below.

\section{Results}
\label{sec:results}
\subsection{Current under electrode level modulation}
\begin{figure*}
    \centering
    \includegraphics[width=\textwidth]{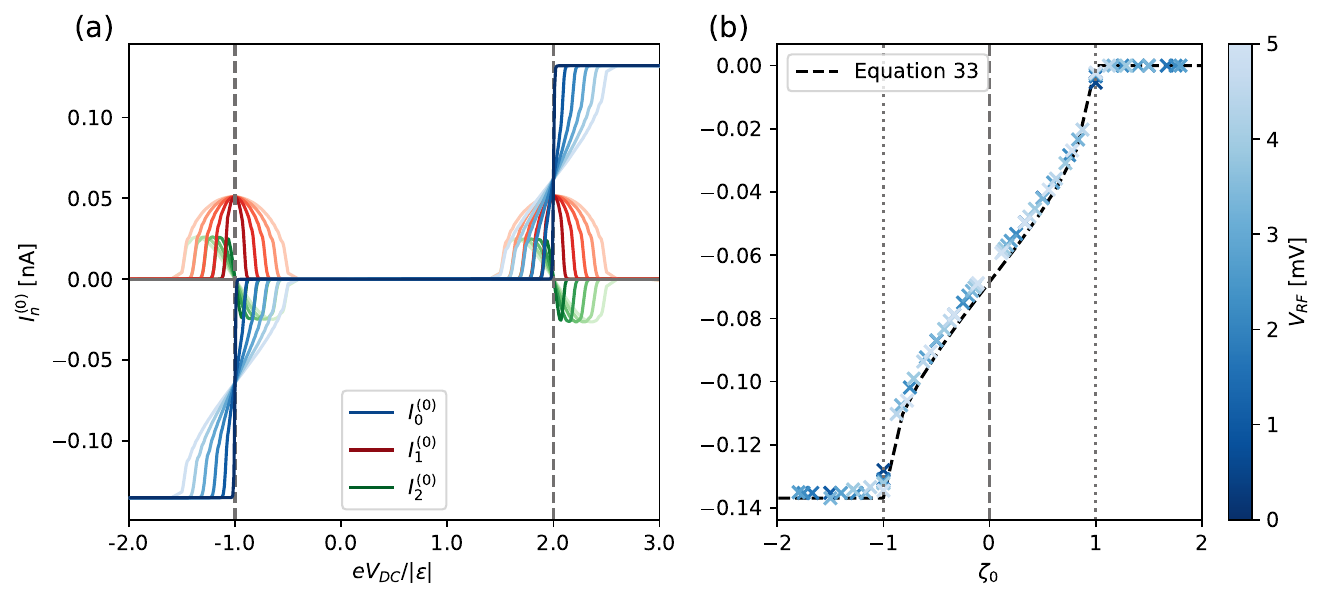}
\caption{\figpanel{a} Unpolarized current components $I_n^{(0)}$ through the QI as a function of the DC bias for different values of $V_{\rm RF}$. The blue curves show the time-averaged component $I_0^{(0)}$, while the red and green curves represent the first and second harmonics, $I_1^{(0)}$ and $I_2^{(0)}$, respectively.
    For each Fourier component, darker (lighter) shades correspond to smaller (larger) values of $V_{\rm RF}$.
    \figpanel{b} Static unpolarized current as a function of $\zeta_0$ for different values of $V_{\rm RF}$ and $V_{\rm DC}$, together with the analytical result in the sharp-threshold limit.
    All parameters not specified in the figure are listed in Table~\ref{tab:app-params}.}
    \label{fig:direct-current}
\end{figure*}

We decompose the current through the impurity into a contribution that survives for an unpolarized electrode, $I_n^{(0)}$, and a spin-sensitive contribution proportional to the electrode polarization, $I_n^{(P)}$, obtaining 
\begin{align}
    I_n = I_n^{(0)} + I_n^{(P)}.
\end{align}
This separation is closely related to the decomposition introduced in Ref.~\onlinecite{Reina-Galvez2025}. We first consider $I^{(0)}$, which isolates the effect of the RF bias on the charge-transport regime, in the limits $\tau_C^{-1}\rightarrow0$ and $T\rightarrow0$. Since cotunneling is not explicitly included in the present theory, the sequential current vanishes in the CB regime and, without RF driving, approaches a step-like function of voltage. When the operating point lies close enough to a QI charge-transition threshold that $\abs{\zeta}<1$, the RF trajectory crosses the threshold during each cycle. The sequential channel is open when the instantaneous voltage, $V(t)$, lies outside the CB region.

Starting from the current expression of Ref.~\onlinecite{Reina-Galvez2025} and retaining the populations of all charge sectors, the unpolarized current at low temperature is
\begin{align}\begin{split}
I_L^{(0)}(t)
=\frac{e\gamma_L}{h}\bigg\{&
\rho_{11}(t)
\left[
\arctan\of{\frac{eV(t)-\varepsilon-U}{\hbar\tau_C^{-1}}}\right.\\
&\hspace{0.7cm}+
\left.\arctan\of{\frac{eV(t)-\varepsilon}{\hbar\tau_C^{-1}}}
\right]
\\
&\hspace{-0.575cm}+
2\rho_{00}(t)
\left[
\frac{\pi}{2}
+
\arctan\of{\frac{eV(t)-\varepsilon}{\hbar\tau_C^{-1}}}
\right]
\\
&\hspace{-0.575cm}-\left.
2\rho_{22}(t)
\left[
\frac{\pi}{2}
-
\arctan\of{\frac{eV(t)-\varepsilon-U}{\hbar\tau_C^{-1}}}
\right]
\right\},
\end{split}
\end{align}
where $V(t)=V_{\rm DC}+V_{\rm RF}\cos\omega t$ and
$\rho_{11}(t)=1-\rho_{00}(t)-\rho_{22}(t)$. The terms proportional to $\rho_{00}(t)$ and $\rho_{22}(t)$ account for processes that begin from the empty and doubly occupied QI, respectively. They are absent from the simplified analytical expression used in Ref.~\onlinecite{Reina-Galvez2025}, where $\rho_{00},\rho_{22}\simeq0$ was assumed. The corresponding polarization-dependent contribution is
\begin{align}\begin{split}
I_L^{(P)}(t)
=-\frac{2e\gamma_L}{h}
P_L
\left\langle{s}_z(t)\right\rangle
\bigg[
&\pi
-\arctan\of{\frac{eV(t)-\varepsilon}{\hbar\tau_C^{-1}}}
\\
+&\arctan\of{\frac{eV(t)-\varepsilon-U}{\hbar\tau_C^{-1}}}
\bigg].
\end{split}
\end{align}

The static component of $I_L^{(0)}$ acquires a simple form in the limits $\tau_C^{-1}\rightarrow0$ and $T\rightarrow0$. As shown in Appendix~\ref{sec:app-regime-switching}, if the charge-state populations remain approximately constant over an RF period, the cycle-averaged current is given by the sequential-current plateau weighted by the fraction of the cycle during which the corresponding sequential channel is open. For the negative-bias transition through the empty state, $\rho_{22}=0$, and
\begin{align}\begin{split}
\label{eq:current-VDC-0}
I^{(0)}_0
\simeq
-\frac{e}{2\hbar}\gamma_L\rho_{11}
\begin{cases}
1, & \zeta_0<-1, \quad \text{Sequential},\\[2pt]
\dfrac{\arccos\of{\zeta_0}}{\pi},
& -1<\zeta_0<1, \quad \text{Mixed},\\[6pt]
0, & \zeta_0>1, \quad \text{CB}.
\end{cases}
\end{split}\end{align}
For the positive-bias transition through the doubly occupied state, $\rho_{00}=0$, and
\begin{align}\begin{split}
\label{eq:current-VDC-2}
I^{(0)}_0
\simeq
\frac{e}{2\hbar}\gamma_L\rho_{11}
\begin{cases}
1, & \zeta_2<-1, \quad \text{Sequential},\\[2pt]
\dfrac{1}{\pi}\arccos\of{\zeta_2},
& -1<\zeta_2<1, \quad \text{Mixed},\\[6pt]
0, & \zeta_2>1, \quad \text{CB}.
\end{cases}
\end{split}\end{align}
The factor $\arccos\of{\zeta_{0,2}}/\pi$ is the fraction of the RF cycle during which the corresponding sequential process is energetically allowed (see Eq~\eqref{eq:seq-frac}).

Figure~\ref{fig:direct-current}\figpanel{a} shows how the different Fourier components of the unpolarized current evolve as the DC operating point is moved through the two charge transitions. The static component $I_0^{(0)}$ approaches the sequential-current plateaus away from the transitions, while the finite RF trajectory smooths the abrupt DC thresholds over the mixed regime. In contrast, the higher harmonics are concentrated around the charge-transition boundaries, where the instantaneous current significantly varies during one RF cycle. Deep inside either the CB or sequential regime, the current is approximately constant along the RF trajectory and the oscillating components are consequently suppressed.

Figure~\ref{fig:direct-current}\figpanel{b} shows the same static current as a function of $\zeta_0$ for different values of $V_{\rm DC}$ and $V_{\rm RF}$. In the sharp-threshold limit, and assuming that the singly occupied population $\rho_{11}$ changes only weakly over the RF cycle, the dependence on $V_{\rm DC}$ and $V_{\rm RF}$ enters mainly through $\zeta_0$. For $-1<\zeta_0<1$, the RF trajectory crosses the charge-transition threshold and the average current is proportional to the fraction of the cycle spent in the sequential regime, $f_{\rm seq} =\arccos(\zeta_0)/\pi$. For $\zeta_0>1$ the trajectory remains in Coulomb blockade, and the sequential current vanishes, whereas for $\zeta_0<-1$ the system remains in the sequential regime throughout the cycle and the current approaches its plateau value. This explains why results obtained for different $V_{\rm DC}$ and $V_{\rm RF}$ follow approximately the same curve when plotted against $\zeta_0$.

\subsection{Spin Dynamics}

Following \cite{Reina-Galvez2025}, we obtain expressions for $\tau_{rel}^{-1}$, the spin-injection term $M(t)$, and the exchange field $B_{xc}(t)$ under RF driving. The oscillating components of $M(t)$ and $B_{xc}(t)$ generate the DLT and FLT contributions, respectively,
\begin{widetext}
    \begin{align}\begin{split}
    \hbar \tau_{rel}^{-1}\of{t}
    =&\ \frac{\gamma_L}{2\pi}
    \ofs{
    \pi
    -\arctan\of{\frac{e\of{V_{\rm DC}+V_{\rm RF}\cos\omega t}-\varepsilon}
    {\hbar\tau_C^{-1}}}
    +\arctan\of{\frac{e\of{V_{\rm DC}+V_{\rm RF}\cos\omega t}-\varepsilon-U}
    {\hbar\tau_C^{-1}}}
    }\\
    &+\frac{\gamma_R}{2\pi}
    \ofs{
    \pi
    +\arctan\of{\frac{\varepsilon}{\hbar\tau_C^{-1}}}
    -\arctan\of{\frac{\varepsilon+U}{\hbar\tau_C^{-1}}}
    },\\
    M\of{t}
    =&\ 
    \frac{\gamma_L}{2\pi}P_L
    \ofs{
    \arctan\of{\frac{e\of{V_{\rm DC}+V_{\rm RF}\cos\omega t}-\varepsilon-U}
    {\hbar\tau_C^{-1}}}
    +
    \arctan\of{\frac{e\of{V_{\rm DC}+V_{\rm RF}\cos\omega t}-\varepsilon}
    {\hbar\tau_C^{-1}}}
    },\\
    g\mu_B B_{xc}\of{t}
    =&\ 
    \frac{\gamma_L}{4\pi}P_L
    \ln\of{
    \frac{
    \of{\varepsilon-e\of{V_{\rm DC}+V_{\rm RF}\cos\omega t}}^2
    +\of{\hbar\tau_C^{-1}}^2
    }{
    \of{\varepsilon+U-e\of{V_{\rm DC}+V_{\rm RF}\cos\omega t}}^2
    +\of{\hbar\tau_C^{-1}}^2
    }} .
\end{split}\end{align}
\end{widetext}

The analytical expression for $M(t)$ above neglects the transient-state populations $\rho_{00}$ and $\rho_{22}$. This approximation is reasonably accurate for the parameters considered here, although deviations can arise near the charge-transition thresholds, where these populations become finite. The numerical calculations therefore use the full population-dependent expression for $M(t)$.

For $\hbar\tau_C^{-1}\ll |\varepsilon|,\ \varepsilon+U$, with $\varepsilon<0<\varepsilon+U$, the right-electrode contribution to $\tau_{rel}^{-1}$ vanishes. If the RF trajectory probes only one charge-transition threshold at a time, the results of Appendix~\ref{sec:app-regime-switching} give
\begin{align}\begin{split}
    \hbar \tau_{rel}^{-1}
    &=
    \frac{\gamma_L}{2}
    \begin{cases}
        1 \quad& \zeta_{0,2}<-1,\\
        \frac{1}{\pi}\arccos\of{\zeta_{0,2}}
        \quad& \abs{\zeta_{0,2}}<1,\\
        0 \quad& \zeta_{0,2}>1.
    \end{cases}
\end{split}\end{align}

\begin{figure*}
    \centering
    \includegraphics[width=\linewidth]{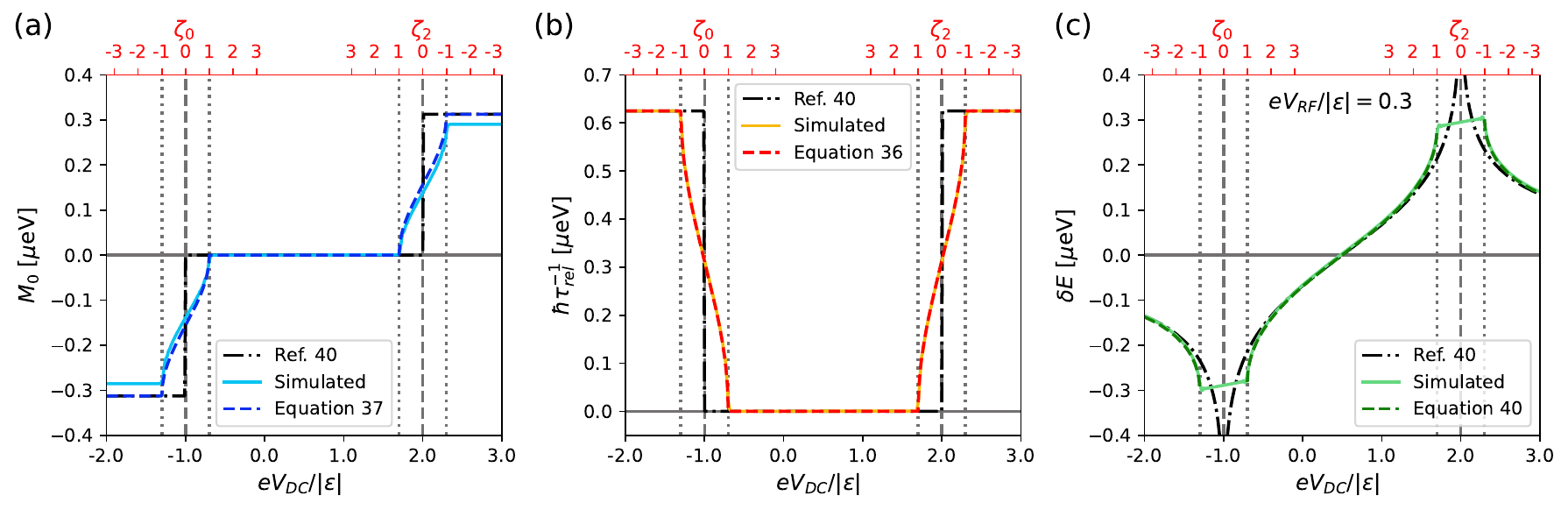}
    \caption{Static spin-dynamical quantities as a function of the DC bias for fixed $V_{\rm RF}/\abs{\varepsilon}=0.3$. 
    \figpanel{a} Static spin accumulation $M_0$. 
    \figpanel{b} Relaxation rate $\tau_{\rm rel}^{-1}$.  \figpanel{c} Lamb shift $\delta E$.
    Solid curves show the full numerical calculation, dashed curves the analytical expressions derived below, and dash-dotted curves the effective barrier-modulation treatment of Ref.~\onlinecite{Reina-Galvez2025}. The upper axes show the corresponding values of $\zeta_0$ and $\zeta_2$. All parameters not specified in the figure are listed in Table~\ref{tab:app-params}.}
    \label{fig:LambSTT0}
\end{figure*}
\begin{figure*}
    \centering
    \includegraphics[width=\linewidth]{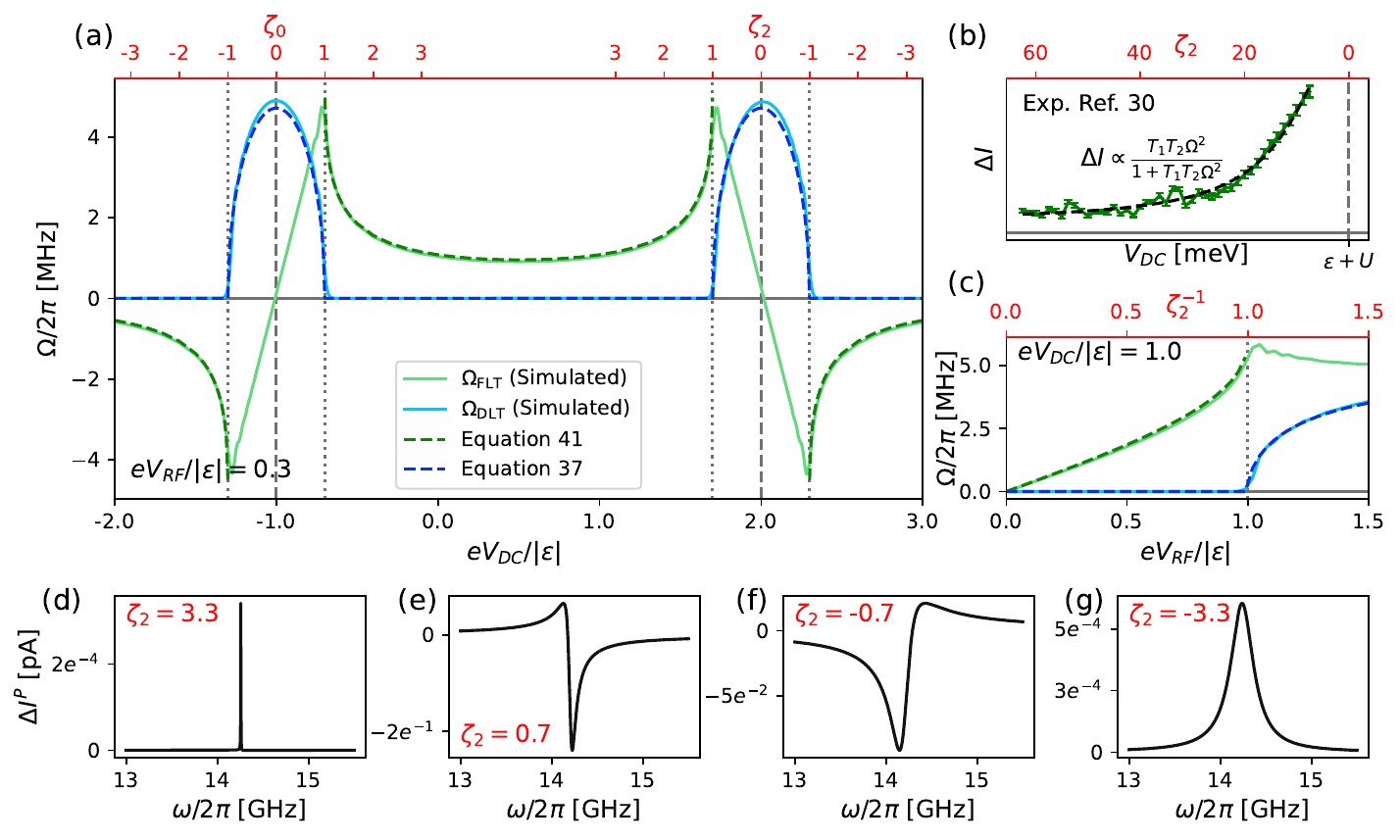}
    \caption{\figpanel{a} Dependence of $\Omega_{DLT}$ and $\Omega_{FLT}$ on $V_{\rm DC}$ for $V_{\rm RF}/\abs{\varepsilon}=0.3$, obtained from the full simulation and from the analytical expressions in Eqs.~\eqref{eq:FLT-expresion} and~\eqref{eq:STT-expresion}. The sign of $\Omega$ specifies the phase of the effective drive and becomes relevant when FLT and DLT coexist. \figpanel{b} Experimental ESR peak amplitude from Ref.~\onlinecite{Greule2026}. The dashed line shows the $\Omega\propto\zeta^{-1}$ relation predicted by our theory fitted to the experimental data. \figpanel{c}, $\Omega_{FLT}$ and $\Omega_{DLT}$ as a function of $V_{\rm RF}$ for fixed $eV_{\rm DC}/\abs{\varepsilon}=1$. \figpanel{d}--\figpanel{g}, Normalized ESR response for representative values of $\zeta_2$ across the CB, mixed, and sequential regimes. 
    The response in the cotunneling regime, \figpanel{d}, is small due to vanishing current, and would likely be enhanced by inclusion of higher-order processes. 
    Meanwhile, in the sequential regime, \figpanel{g}, the signal vanishes due to short decoherence times. 
    All parameters not specified in the figure are listed in Table~\ref{tab:app-params}.}
    \label{fig:Rabi-and-esr}
\end{figure*}

The static and oscillatory components of the spin-injection term can be written as
\begin{align}\begin{split}
\label{eq:STT-expresion}
    M\of{t}
    &=M_0+\sum_{n=1}^{\infty}M_n\cos n\omega t\\
    M_0
    &=
    \frac{\gamma_L}{2}P_L\text{sign}\of{V_{\rm DC}}
    \begin{cases}
        1 \quad& \zeta_{0,2}<-1,\\
        \dfrac{\arccos\of{\zeta_{0,2}}}{\pi}
        \quad& -1<\zeta_{0,2}<1,\\
        0 \quad& \zeta_{0,2}>1,
    \end{cases}\\
    M_n
    &=
    \frac{\gamma_L}{\pi n}P_L
    \begin{cases}
        \sin\ofs{n\arccos\of{-\zeta_0}}
        \quad& -1<\zeta_0<1,\\
        \sin\ofs{n\arccos\of{\zeta_2}}
        \quad& -1<\zeta_2<1,\\
        0 \quad& \text{otherwise}.
    \end{cases}
\end{split}\end{align}

Both $\tau_{rel}^{-1}$ and $M(t)$ originate from real charge-transfer processes and therefore display the same threshold structure as the transport current. The relaxation rate is positive by construction, whereas the sign of $M(t)$ follows the direction of spin injection.
Fig.~\ref{fig:LambSTT0}\figpanel{a} and \figpanel{b} show $\tau_{rel}^{-1}$ and the static part, $M_0$, as a function of the DC bias.
Within the analytical approximation above, both quantities are finite in the sequential regime and vanish deep inside the CB regime.
The present treatment replaces the abrupt step-function threshold from Ref.~\onlinecite{Reina-Galvez2025} by the smooth $\arccos\of{\zeta}$ dependence.
Cotunneling would generate additional contributions to DLT and $\tau_{rel}^{-1}$ in the CB regime, which are missing from the present master equation.

Fig.~\ref{fig:LambSTT0}\figpanel{a} and \figpanel{b} highlight the common origin of $M_0$ and $\tau_{\rm rel}^{-1}$ in real charge-transfer processes. Both are strongly suppressed inside the CB region and approach approximately constant values in the sequential regime, while the finite RF trajectory broadens the crossover over the mixed regime. The sign of $M_0$ reverses between the two bias directions because the direction of spin injection changes, whereas $\tau_{\rm rel}^{-1}$ remains positive because tunneling relaxes the QI spin independently of the direction of charge transfer.

We next consider the dependence of the exchange magnetic field on $V_{\rm RF}$ and $V_{\rm DC}$.
When $\tau_C^{-1}$ is small compared with the distance to the nearest QI energy, the result can be written as (see Appendix~\ref{sec:app-exch-four})
\begin{align}\begin{split}
\label{eq:exchange-field}
    g\mu_B &B_{xc}\of{t}
    =\
    \frac{\gamma_L}{2\pi}P_L
    \left\{
    \ln\abs{\frac{\varepsilon-eV_{\rm DC}}
    {\varepsilon+U-eV_{\rm DC}}}\right.\\
    &\left.+
    \ln\frac{1+r_2^2}{1+r_0^2}
    +2\sum_{n=1}^{\infty}
    \frac{r_2^n-r_0^n}{n}\cos (n\omega t)\right\} .
\end{split}\end{align}
Here,
\begin{align}\begin{split}
    r_0
    &=
    -\frac{\text{sign}\of{\zeta_0}}
    {\abs{\zeta_0}+\sqrt{\zeta_0^2-1}},
    \qquad
    r_2
    =
    \frac{\text{sign}\of{\zeta_2}}
    {\abs{\zeta_2}+\sqrt{\zeta_2^2-1}},
\end{split}\end{align}
and for small $V_{\rm RF}$,
$r_0\approx-1/(2\zeta_0)$ and $r_2\approx1/(2\zeta_2)$.
This expression applies only for $\abs{\zeta_{0,2}}>1$. When the RF trajectory reaches a singular point, the finite broadening $\tau_C^{-1}$ must be restored.

The first term of Eq.~\eqref{eq:exchange-field} is the zero-RF exchange field, whose projection along the spin direction gives the theoretically and experimentally established Lamb shift~\cite{Reina-Galvez2025,Greule2026}.
The $V_{\rm RF}$-dependent correction changes its magnitude only weakly as long as the RF trajectory remains sufficiently far from the charge-transition thresholds,

\begin{align}\begin{split}
\label{eq:lamb-shift}
    \delta E
    =&\
    g\mu_B\cos\theta B_{xc;0}\\
    = &\
    \frac{\gamma_L}{2\pi}P_L\cos\theta
    \ofs{
    \ln\abs{\frac{\varepsilon-eV_{\rm DC}}
    {\varepsilon+U-eV_{\rm DC}}}
    +
    \ln\frac{1+r_2^2}{1+r_0^2}
    }\\
    \approx&\
    \frac{\gamma_L}{2\pi}P_L\cos\theta
    \Bigg[
    \ln\abs{\frac{\varepsilon-eV_{\rm DC}}
    {\varepsilon+U-eV_{\rm DC}}}\\
    &+
    \frac{\of{eV_{\rm RF}}^2}{4}
    \of{
    \frac{1}{\of{\varepsilon+U-eV_{\rm DC}}^2}
    -
    \frac{1}{\of{\varepsilon-eV_{\rm DC}}^2}
    }
    \Bigg]\\
    &+    \mathcal{O}\of{V_{\rm RF}^4}.
\end{split}\end{align}
As illustrated in Fig.~\ref{fig:LambSTT0}\figpanel{c}, the static exchange-field contribution, $B_{{\rm xc};0}$, and hence the Lamb shift $\delta E=g\mu_B B_{{\rm xc};0}\cos\theta$, remain close to the result of the effective tunneling-modulation model of Ref.~\onlinecite{Reina-Galvez2025} over most of the CB regime. However, the two descriptions differ in how the RF drive enters the static component. In the tunneling-modulation model, $\Gamma_L(t)\propto(1+A\cos\omega t)^2$, the RF drive primarily modulates the magnitude of the coupling while the energy detuning entering the exchange field remains fixed by $V_{\rm DC}$. Its time-averaged component therefore largely preserves the DC-bias dependence of the static exchange field.

In the present treatment, by contrast, the RF voltage enters directly in the energy arguments, $B_{\rm xc}(t)=B_{\rm xc}[V_{\rm DC}+V_{\rm RF}\cos\omega t]$. Consequently, $B_{{\rm xc};0}$ corresponds to an average of the exchange field over the voltage interval explored during the RF cycle. Deep inside the CB regime, where $B_{\rm xc}$ varies slowly with bias, this produces only a small correction and the two approaches give similar results. Close to a charge-transition threshold, the exchange field varies strongly with bias, and this averaging substantially modifies and smooths the static response. The approximately linear dependence near the center of the CB region, together with the pronounced increase in magnitude near the charge-transition thresholds, has also been observed experimentally~\cite{Greule2026}.

In the zero-temperature and zero-broadening limit, the static expression of Ref.~\onlinecite{Reina-Galvez2025} diverges at a charge-transition threshold. This is an artifact of the sharp-level approximation rather than a physical divergence of the resonance frequency. A finite RF trajectory averages the exchange field over a range of biases and therefore smooths its static component in the mixed regime; although a finite broadening is still required for numerical reasons, its influence on the dynamics is substantially reduced.

Taken together, the three panels of Fig.~\ref{fig:LambSTT0} emphasize the distinction between real and virtual charge-transfer processes. The nonequilibrium spin accumulation $M_0$ and the relaxation rate $\tau_{\rm rel}^{-1}$ follow the opening and closing of real sequential channels and are therefore strongly suppressed in CB. The exchange-field contribution behaves differently: it remains finite throughout the CB region and varies continuously with $V_{\rm DC}$ because it originates from virtual charge fluctuations. These different static behaviors anticipate the first-harmonic response of Fig.~\ref{fig:Rabi-and-esr}: DLT is localized near the charge transitions, whereas FLT remains finite farther inside the CB regime.

The remaining cosine coefficients of the exchange field are
\begin{align}\begin{split}
\label{eq:FLT-expresion}
    g\mu_B B_{xc,n}
    =&\
    \frac{\gamma_LP_L}{n\pi}
    \of{r_2^n-r_0^n}\approx 
    \frac{\gamma_LP_L\of{eV_{\rm RF}}^n}
    {2^n n\pi}\\
    &\times \ofs{
    \frac{1}{\of{\varepsilon+U-eV_{\rm DC}}^n}
    -
    \frac{1}{\of{\varepsilon-eV_{\rm DC}}^n}
    }\\
    &+
    \mathcal{O}\of{V_{\rm RF}^{n+2}}.
\end{split}\end{align}
For the first harmonic and weak RF bias, the same result may be written as
\begin{align}
g\mu_B B_{xc,1}
\simeq
eV_{\rm RF}
\frac{\partial}{\partial\of{eV_{\rm DC}}}
\left[g\mu_B B_{xc,0}\right],
\end{align}
which shows that the FLT drive is controlled by the bias derivative of the static exchange field rather than by its absolute magnitude.

The FLT Rabi frequency associated with first-harmonic driving is
\begin{align}\begin{split}
    \Omega_{FLT}
    =\frac{g\mu_B}{2\hbar}B_{xc;1}\sin\theta ,
\end{split}\end{align}
whereas the DLT Rabi frequency is
\begin{align}\begin{split}
    \Omega_{DLT}
    = \frac{1}{2\hbar}M_1\sin\theta.
\end{split}\end{align}
Both quantities are plotted as functions of $V_{\rm DC}$ and $V_{\rm RF}$ in Fig.~\ref{fig:Rabi-and-esr}\figpanel{a} and \figpanel{c}, together with the corresponding analytical results from Eqs.~\eqref{eq:FLT-expresion} and~\eqref{eq:STT-expresion}. 
The field-like contribution $\Omega_{FLT}$ characterizes coherent field-driven spin dynamics, whereas $\Omega_{DLT}$ quantifies the strength of the dissipative spin-injection drive. 
Combined they lead to an effective Rabi frequency $\Omega = \Omega_{DLT}+ i\Omega_{STT}$.
When both mechanisms are present, their interference depends on their relative phase and on the dynamical response of the spin. 
This distinction is important for interpreting the asymmetric ESR line shapes discussed below.

In Fig.~\ref{fig:Rabi-and-esr}\figpanel{a}, the DLT is concentrated near the charge-transition thresholds, where $\abs{\zeta_{0,2}}<1$ and the RF trajectory periodically crosses between the CB and sequential regimes, while $M_1$ rapidly vanishes away from these regions. This differs from Ref.~\onlinecite{Reina-Galvez2025}, where $\Omega_{DLT}$ remains finite throughout the sequential regime. The FLT, by contrast, persists throughout the CB regime, depends only weakly on $V_{\rm DC}$ far from the thresholds, and becomes strongly enhanced as a transition is approached. Figure~\ref{fig:Rabi-and-esr}\figpanel{b} provides a qualitative comparison with the experimental bias dependence reported in Ref.~\onlinecite{Greule2026}. The inferred Rabi frequency varies only weakly with $V_{\rm DC}$ inside the CB regime and increases strongly on approaching the charge-transition threshold, consistent with the calculated FLT behavior in panel a.

The $V_{\rm RF}$ sweep in Fig.~\ref{fig:Rabi-and-esr}\figpanel{c} shows the same crossover from a different perspective. At small $V_{\rm RF}$, the system remains within the CB regime and $\Omega_{FLT}$ increases approximately linearly with $V_{\rm RF}$. As the RF trajectory approaches $\abs{\zeta_2}=1$, the FLT response departs from this linear behavior and develops a broad plateau, while the DLT appears as real charge-transfer processes begin to be modulated. Increasing $V_{\rm RF}$ therefore changes both the overall driving strength and the relative importance of FLT and DLT. This behavior follows naturally from treating the RF bias explicitly. Varying $V_{\rm DC}$ shifts the center of the RF trajectory through the transport landscape, whereas increasing $V_{\rm RF}$ expands the range explored during each cycle.

The ESR spectra in Fig.~\ref{fig:Rabi-and-esr}\figpanel{d}-\figpanel{g} show the corresponding evolution of the resonance line shape. 
In line with previous works we write the ESR response as
\begin{equation}
        \Delta I^{(P)} 
        =   \Delta I_s \frac{\Gamma }{\Gamma^2+\of{\omega-\omega_0}^2} 
        +   \Delta I_a \frac{\omega- \omega_0}{\Gamma^2 + \of{\omega-\omega_0}^2},
\end{equation}
where $\omega_0$ is the resonance frequency $\Gamma$ is the line-width, $\Delta I_{s(a)}$ is the magnitude of (a)symetric response.
Away from the mixed regime, the resonance is approximately symmetric, whereas for $\abs{\zeta_2}<1$ the simultaneous presence of DLT and FLT produces a strongly asymmetric response due to their different dynamical phases relative to the RF drive, in line with theoretical analysis from Ref. \cite{Ruckert2026}. Similar behavior has been observed experimentally~\cite{Kovarik2024}. A detailed analysis of the dynamical phases acquired by the DLT- and FLT-induced spin responses relative to the RF driving voltage, and of the resulting interference between the two mechanisms, is left for future work.

\section{Conclusion}
We studied the RF bias as a driving mechanism in ESR-STM by employing a Redfield QME description of a generalized Anderson impurity with an oscillating voltage directly in the electronic reservoirs. 
After a gauge transformation, the drive appears through Bessel-weighted Floquet sidebands in the tunneling rates.
In the semiclassical regime $eV_{\rm RF}\gg\hbar\omega$, this Floquet structure reduces to an instantaneous-bias description in which the system samples the static transport response along $V(t)=V_{\rm DC}+V_{\rm RF}\cos\omega t$.

The RF bias has two main effects.
Firstly, it averages the transport rates over its bias window, alternating the magnitude of the static part of the rates.
Secondly, it induces an oscillation of these rates, thus driving the ESR. 
Both effects are dramatically amplified when the RF bias is sufficiently strong to cross the threshold energy between CB and sequential regions.
When no such crossing occurs, the static rates agree with those predicted by previous works \cite{Reina-Galvez2025}.

The oscillatory components of the transition rates, FLT and DLT, show a novel behaviour. 
Unlike previous works, we predict that DLT only appears near the threshold energies. Meanwhile, we show that far from the threshold, FLT is inversely proportional to the difference between the applied voltage and the threshold energies, consistent with the experimental trend reported in Ref.~\onlinecite{Greule2026}.
The complex interplay between DLT and FLT dynamics near the impurity threshold creates a highly asymetric ESR peak, in agreement with experimental observations from Ref.~\onlinecite{Kovarik2024}.

The presented work strongly supports the previously proposed exchange field-mediated driving mechanism in ESR-STM~\cite{Reina-Galvez2019,Kovarik2024,Reina-Galvez2025,Zhang2026_ACS_Nano}. 
We clarify the microscopic processes leading to oscillations in hopping rates and show that the resulting dynamical observables match experimental trends. 
However, a quantitative description of the measurable ESR signal deep in the CB regime requires including higher-order cotunneling processes.
Furthermore, our work shows that additional oscillations in the hopping rates can be induced by including bias-dependent hopping amplitude and density of states. 
The investigation of these phenomena is left for future work. 

\begin{acknowledgments}
J. R.-G. acknowledges support by the German Research Foundation [Deutsche Forschungsgemeinschaft (DFG)] under Projects No.~450396347 and No.~425217212 - SFB 1432. C. W. acknowledges support from the Institute for Basic Science (IBS-R027-D1). J. R.-G. acknowledges insightful discussions with H. T. Bui during the preparation of this manuscript. M. N. and N. L. Acknowledge funding from estrategia IKUR, the ``Grupos Consolidados'' project IT1837-26, Department of Education and Research, both from the Basque Government and the project PID2021-127917NB-I00 funded by MCIN/AEI/ 10.13039/501100011033. M. N. acknowledges insightful discussions with S. R. McMillan.
\end{acknowledgments}

\bibliographystyle{ieeetr}
\bibliography{ref}
\onecolumngrid

\appendix
\section{Derivation of QME}
\label{sec:app-rates}
Here we derive the master equation governing $\rho_{ij}\of{t}$.
We start with the Heisenberg equation of motion 
\begin{align}\begin{split}
    \partial_t \rho_{lj}\of{t} 
    &= \frac{i}{\hbar} \braket{\ofs{\hat{H}, \hat{\rho}_{lj}}} 
    = \frac{i}{\hbar} \braket{\ofs{\hat{H}_{QI}, \hat{\rho}_{lj}}} 
    + \frac{i}{\hbar} \braket{\ofs{\hat{H}_{T}, \hat{\rho}_{lj}}} \\
    &= \frac{i}{\hbar} \Delta_{lj} + \sum_{\alpha k \sigma uv} 
        t_\alpha^0 \lambda_{ul\sigma} \braket{c_{\alpha k\sigma}^{\dag} \hat{\rho}_{uj}} \sum_n J_{\alpha n} e^{in\omega t}
    -   t_\alpha^0 \lambda_{ju\sigma} \braket{c_{\alpha k\sigma}^{\dag} \hat{\rho}_{lu}} \sum_n J_{\alpha n} e^{in\omega t} \\
    &+   t_\alpha^{0} \lambda_{lu\sigma}^{*} \braket{\hat{\rho}_{uj} c_{\alpha k\sigma}} \sum_n J_{\alpha n} e^{-in\omega t}
    +   t_\alpha^{0} \lambda_{uj\sigma}^{*} \braket{\hat{\rho}_{lu} c_{\alpha k\sigma}} \sum_n J_{\alpha n} e^{-in\omega t} 
\end{split}\end{align}
where $J_{\alpha n} = J_n \of{\frac{eV^{\rm RF}_\alpha}{\hbar \omega}}$. 

To proceed, we need to evaluate the mixed Green's functions $\braket{\hat{\rho}_{lj} c_{\alpha k\sigma}}$ and $\braket{c_{\alpha k\sigma}^{\dag} \hat{\rho}_{lj}}$ by applying the Langreth theorem to the first order of perturbation theory \cite{Rammer2007}.
We refer to the derivation in Ref~\cite{Reina-Galvez2021}, and we arrive at the QME in Fourier space
\begin{align}\begin{split}
    \of{n\hbar\omega+ \Delta_{lj}} \rho_{lj;n}
    = i \sum_{vu;m}& \big[\Gamma_{vl,ju;m}\rho_{vu;n-m} - \Gamma^{*}_{lv,vu;-m}\rho_{uj;n-m} \\
    -&   \Gamma_{jv,vu;m}\rho_{lu;n-m} + \Gamma^*_{vj,lu;-m}\rho_{uv;n-m} \big ] 
\end{split}\end{align} 
The rates $\Gamma_{vl,ju,m}$ are given as 
\begin{align}\begin{split}
\label{eq:app-rates}
    \Gamma_{vl,ju;n} = \frac{1}{2\pi \hbar} &\sum_{\alpha\sigma;m} \lambda_{vl\sigma}\lambda_{uj\sigma}^* \int_{-\infty}^{\infty} d\epsilon 
    \frac{\Sigma_{\alpha\sigma;m,m-n}^{<} \of{\epsilon -n\hbar\omega}}{\epsilon - \Delta_{ju} +i\hbar\tau_C^{-1}} \\
    +& \sum_{\alpha\sigma;m} \lambda_{lv\sigma}^* \lambda_{ju\sigma} 
    \int_{-\infty}^{\infty} d\epsilon \ofs{2\pi i \delta\of{\epsilon - \Delta_{uj}} \Sigma_{\alpha\sigma;m+n,n}^r\of{\epsilon - n\hbar\omega}
    + \frac{\Sigma_{\alpha\sigma;m+n,m}\of{\epsilon-n\hbar\omega}}{\epsilon-\Delta_{uj} -i\hbar\tau_C^{-1}}}
\end{split}\end{align}
where $\Sigma_{\alpha\sigma;nm}^{C}\of{\epsilon}$ are the Fourier-Floquet components of the self-energy of the reservoir. 
Here we substitute the effective time dependence from Ref.~\onlinecite{Reina-Galvez2021} introduced via the parameter $A_\alpha$ with the Bessel function time dependence originating from the RF bias $t_\alpha^0\of{1+A_\alpha\cos\of{\omega t}} \rightarrow t_\alpha^0 \sum_n J_{n\alpha} e^{in\omega t}$. 
\begin{align}\begin{split}
\label{eq:app-self-energy}
    \Sigma_{\alpha\sigma;nn'}^{r,a}\of{\epsilon} = \frac{1}{\hbar} \abs{t_\alpha^0}^2
        \sum_m J_{\alpha;n-m} J_{\alpha;n'-m}  \int d \xi_{\alpha k\sigma}  
        \frac{D_\alpha\of{\xi_{\alpha k \sigma}}}{\epsilon + m\omega -\xi_{\alpha k \sigma} \pm i\hbar\tau_C^{-1}} \\
    \Sigma_{\alpha\sigma;nn'}^{<}\of{\epsilon} = \frac{2\pi i}{\hbar} \abs{t_\alpha^0}^2
        \sum_m J_{\alpha;n-m} J_{\alpha;n'-m} f^{+}\of{\epsilon+m\omega} D_\alpha\of{\epsilon+m\omega}
\end{split}\end{align}
Finally, substituting equation~\ref{eq:app-self-energy} into equation~\ref{eq:app-rates}, we obtain equation~\ref{eq:rate-four}.

In the time-independent case, the resulting QME for a two-level system \cite{Reina-Galvez2025} can be mapped to \cite{Braun2004}. 
\begin{align}\begin{split}
    \Gamma_{\alpha} n_\alpha^{F-}\of{\varepsilon} &\rightarrow 2\Re\ofc{\Gamma_{g00g}^0\of{t}}\\
    \Gamma_{\alpha} n_\alpha^{F-}\of{\varepsilon+U} &\rightarrow 2\Re\ofc{\Gamma_{2gg2}^0\of{t}}\\
    \Gamma_{\alpha} n_\alpha^{F+}\of{\varepsilon} &\rightarrow 2\Re\ofc{\Gamma_{0gg0}^0\of{t}}\\
    \Gamma_{\alpha} n_\alpha^{F+}\of{\varepsilon+U} &\rightarrow 2\Re\ofc{\Gamma_{g22g}^0\of{t}}\\
\end{split}\end{align}
In the theory from Ref.~\onlinecite{Braun2004}, RHS, $\Gamma_\alpha$ refers to the strength of the coupling of electrode $\alpha$ to the QD. 
The main additions of Ref.~\onlinecite{Reina-Galvez2021} are the finite broadening $\tau_C^{-1}$, harmonic time dependence, and a generalization to a more complex spin Hamiltonian.

\section{Semiclassical Approximation to Floquet Bands}
\label{sec:rem-bes-fun}
\newcommand{\Jsq}[1]{\ensuremath{J_{#1}^2\of{z}}}
\newcommand{\J}[1]{\ensuremath{J_{#1}\of{z}}}
Consider the following identity including the Bessel functions of the first kind $\J{n}$
\begin{align}\begin{split}
    \sum_{n} \J{n+m}\J{n} f(x_0+an) 
\end{split}\end{align}
where  $f\of{x}$ is a continuous and infinitely differentiable function around $x_0$. 
Taylor expanding $f\of{x}$ around $x_0$, we obtain the following.
\begin{align}
     \sum_{n} \J{n+m}\J{n} f(x_0+an) = \sum_{n=0} \frac{a^k f^{\of{k}}(x_0)}{k!} \sum_n \J{n+m}\J{n} n^k 
\end{align}
Next, we use the following identity
\begin{align}\begin{split}
    \frac{2n}{z} J_n = \J{n-1} + \J{n+1}
\end{split}\end{align}
and write down the expression for the first couple of $k$
\begin{align}\begin{split}
    \sum_{n} \J{n+m}\J{n} &= \delta_{m0}\\
    \sum_{n} \J{n+m}\J{n} n &= \of{\frac{z}{2}}^1 \sum_n \J{n+m}\of{\J{n-1} + \J{n+1}}\\
    \sum_{n} \J{n+m}\J{n} n^2 &= \of{\frac{z}{2}}^2 \sum_n \J{n+m} \of{\J{n-2}+\J{n}}\frac{n}{n-1} +\of{ \J{n} + \J{n+2}}\frac{n}{n+1} \\
    \sum_{n} \J{n+m}\J{n} n^3 &= \of{\frac{z}{2}}^3 \sum_n \J{n+m} \big(\of{\J{n-3}+\J{n-1}}\frac{n^2}{\of{n-1}\of{n-2}} + \\
    &+ \of{\J{n-1} + \J{n+1}}\frac{n^2}{n^2-1}
    \of{\J{n+1} + \J{n-3}}\frac{n}{\of{n+1}\of{n+2}}\big) \\
\end{split}\end{align}
In general, we can write 
\begin{align}\begin{split}
    \sum_n \J{n+m}\J{n} n^k = \of{\frac{z}{2}}^k C_k^m
\end{split}
\end{align}
where $C_k^m$ is given by a sum over Bessel functions as the ones above. 

Now let us assume $z\gg a$. 
Small $a$ allows us to consider only moderate values of $k$. 
Furthermore, a large $z$ implies that only Bessel functions of large order will contribute. 
This implies that only terms of the sum that contribute will satisfy $n\gg k$.
As a result, we can estimate the fractions involving $n$ in $C_k$ as 1. 

\begin{align}\begin{split}
    C_0^m &= \sum_n \J{n+m}\J{n} \\
    C_2^m &= \sum_n \J{n+m} \of{\J{n-1} + \J{n+1}} \\
    C_4^m &\approx \sum_n \J{n+m} \of{\J{n-2} + 2\J{n} + \J{n+2}}  \\
    C_6^m &\approx \sum_n \J{n+m} \of{\J{n-3}+3 \J{n-1} + 3 \J{n+1} + \J{n+3}} \\
\end{split}\end{align}
Considering the term inside the square, we notice that increasing $k \rightarrow n+1$ affects one of the Bessel functions $\J{n}$ as $\J{n}  \rightarrow \J{n+1} + \J{n-1}$. 
Furthermore, the orthogonality of the Bessel function leads to $\sum_n \J{n+m} \J{n+m'} = \delta_{mm'}$. 
An analogous behavior can be obtained by integrating harmonic exponentials. 
\begin{align}\begin{split}
    C_k^m \approx \frac{1}{2\pi}  \int_0^{2\pi} e^{iym}\of{e^{iy} + e^{-iy}}^k dy =  \frac{2^k}{2\pi} \int_0^{2\pi} e^{iym} \cos^k\of{y} dy
\end{split}\end{align}
For $\of{e^{iy} + e^{-iy}}^{k}$ we obtain the same behavior upon increasing $k\rightarrow k+1$: $e^{iky} \rightarrow e^{i\of{k+1}y} + e^{i\of{k-1}y}$, while the integral reproduces the orthogonal behavior. 

\begin{align}\begin{split}
     \sum_{n} \J{n+m}\J{n} f(x_0+an)
     &= \frac{1}{2\pi} \sum_{k=0} \frac{\of{az}^k f^{\of{k}}(x_0)}{k!}  \int_0^{2\pi} e^{iym} \cos^k\of{y} dy \\
     &= \frac{1}{2\pi} \int_0^{2\pi} e^{iym} \sum_{k=0} \frac{\of{az\cos\of{y}}^k f^{\of{k}}(x_0)}{k!}    dy \\
     &= \frac{1}{2\pi}\int_0^{2\pi} e^{iym} f\of{x_0 + az\cos\of{y}}dy \\
\end{split}\end{align}
Thus, we can make the following statement. 
For $z\gg1$, and $f\of{x}$ being a smooth function for $x\in (x_0-a, x_0+a)$ 
\begin{align}\begin{split}
    f\of{x_0 + a\cos y} = \sum_{nm} J_{n+m}\of{z}J_{m}\of{z}  f\of{x_o + \frac{a}{z}m} e^{-iny}
\end{split}\end{align}

Applying this to our system, we have the $m$-th components of the Fourier expansion written as 
\begin{align}\begin{split}
    \label{eq:app-remove-bessel}
    \sum_n J_{n}\of{\frac{eV^{RF}_\alpha}{\hbar\omega}} J_{n+m}\of{\frac{eV^{RF}_\alpha}{\hbar\omega}} I^{\pm}\of{E \pm \hbar n\omega} \approx  \frac{1}{2 \pi} \int_0^{2\pi} e^{i t}I^{\pm} \of{E\pm eV^{RF}_\alpha\cos t} dt
\end{split}
\end{align}
for $eV^{RF}_\alpha \gg \hbar \omega$.
We notice that RHS is the Fourier component of $I^{\pm}\of{E+eV^{RF}_\alpha\cos\of{\omega t}}$, for an arbitrary $\omega$.
As a result, we can write that upon applying a time-dependent bias 
\begin{align}\begin{split}
    V_{DC} &\rightarrow V_{DC} +V^{RF}_\alpha\cos{\omega t} \\
    I^{\pm}(V_{DC}) &\rightarrow I^{\pm}(V_{DC} +V^{RF}_\alpha\cos{\omega t})
\end{split}\end{align}

\section{Charge Transfer Under Oscillating Bias}
\label{sec:app-regime-switching}

\newcommand{\efermi}{\ensuremath{E^{Fermi}_L}}
\newcommand{\Iback}{\ensuremath{I_{DC}^{Seq}\of{0,0}}}
Multiple descriptors of the dynamics depend on the following identity 
\begin{align}\begin{split}
    \arctan\of{\frac{E_{0,2} + eV_{\rm RF}\cos\omega t }{\hbar\tau_C^{-1}}} 
\end{split}\end{align}
where is $E_{0} = V_{\rm DC} - \varepsilon$, and $E_{2} = eV_{\rm DC} - \varepsilon- U$. 

In the limit of $\hbar\tau_C^{-1}\ll \abs{E_{0,2}}$ we can write the $\arctan(E)$ function as a step function $\Theta\of{E}$. 
We then want to find the Fourier expansion of the function
\begin{align}\begin{split}
    \lim_{\hbar\tau_C^{-1}\rightarrow 0} \arctan\of{\frac{E_{0,2} + eV_{\rm RF}\cos\omega t }{\hbar\tau_C^{-1}}} 
    =\pi\of{\Theta\of{E_{0,2} + V_{\rm RF}\cos\omega t } - \frac{1}{2}}= \pi\of{\Theta\of{\pm \zeta_{0,2} + \cos\omega t}  - \frac{1}{2}}
\end{split}\end{align}
where definition $\zeta_{0,2}=\pm \frac{E}{V_{\rm RF}}$ is consistent with Eq.~\eqref{eq:zeta-def}.
$\pm$ refers to $0$ and $2$ respectivelly. 
We can write 
\begin{align}\begin{split}
    \frac{1}{T}\int_0^T \Theta\of{\pm \zeta + \cos \omega t} dt
    = \begin{cases}
        0 \quad& \pm \zeta < -1 \\
        \frac{\arccos \mp \zeta }{\pi} \quad& -1 \leq \pm\zeta < 1\\
        1 \quad& \pm \zeta \geq 1\\
    \end{cases}
\end{split}\end{align}
Thus, we obtain the $0$-th Fourier component of the arctangents. 
\begin{align}\begin{split}
    \lim_{\hbar\tau_C^{-1}\rightarrow 0} \frac{1}{T}\int_0^T \arctan\of{\frac{E_{0,2} + eV_{\rm RF}\cos\omega t }{\hbar\tau_C^{-1}}} dt 
    =\begin{cases}
        -\frac{\pi}{2} \quad& \pm \zeta < -1 \\
        \mp \of{\arccos \zeta - \frac{\pi}{2}} \quad& -1 \leq \pm\zeta < 1\\
        \frac{\pi}{2} \quad& \pm \zeta \geq 1\\
    \end{cases}
\end{split}\end{align}
so that 
\begin{align}\begin{split}
    \lim_{\tau_C^{-1}\rightarrow 0} \frac{1}{T}\int_0^T \ofs{
        \arctan\of{\frac{E_2 + eV_{\rm RF}\cos\omega t }{\hbar\tau_C^{-1}}}
    -   \arctan\of{\frac{E_0 + eV_{\rm RF}\cos\omega t }{\hbar\tau_C^{-1}}}} dt 
    =&  \begin{cases}
        0 \quad&  \zeta_{0} < -1, \zeta_2 \geq 1\\
       \arccos \zeta_0 -\pi \quad& -1 \leq \pm\zeta_0 < 1, \zeta_2 \geq 1\\
       -\pi \quad&  \zeta_0 \geq 1, \zeta_2 \geq 1 \\
       \arccos \zeta_2 -\pi \quad& \zeta_0 \geq 1, -1 \leq \zeta_2 < 1\\
       0 \quad& \zeta_0 \geq 1, \zeta_2 < -1
    \end{cases} \\
     \lim_{\tau_C^{-1}\rightarrow 0} \frac{1}{T}\int_0^T \ofs{
        \arctan\of{\frac{E_2 + eV_{\rm RF}\cos\omega t }{\hbar\tau_C^{-1}}}
     +  \arctan\of{\frac{E_0 + eV_{\rm RF}\cos\omega t }{\hbar\tau_C^{-1}}} }dt 
    =&  \begin{cases}
        -\pi \quad&  \zeta_{0} < -1, \zeta_2 \geq 1\\
       -\arccos \zeta_0 \quad& -1 \leq \pm\zeta_0 < 1, \zeta_2 \geq 1\\
       0 \quad&  \zeta_0 \geq 1, \zeta_2 \geq 1 \\
       \arccos \zeta_2 \quad& \zeta_0 \geq 1, -1 \leq \zeta_2 < 1\\
       \pi \quad& \zeta_0 \geq 1, \zeta_2 < -1
    \end{cases} \\
\end{split}\end{align}
Likewise, we want to evaluate higher Fourier components 
\begin{align}\begin{split}
 \frac{2}{T}\int_0^T \Theta\of{\pm \zeta + \cos (\omega t)} \cos (n\omega t) dt
    &= \frac{1}{\pi} \int_{-\arccos\of{\mp\zeta}}^{\arccos\of{\mp\zeta}} \cos (nx) dx \\
    &=\begin{cases}
         \frac{2}{n\pi}\sin  (n\arccos\of{\mp\zeta}) \quad& -1 \leq \pm\zeta < 1\\
         0\quad& \text{otherwise}\\
    \end{cases}
\end{split}\end{align}

\section{Exchange Magnetic Field Under Oscillating Bias}
\label{sec:app-exch-four}
For small $\tau_C^{-1}$ and $T$ we can write the exchange magnetic field as
\begin{align}\begin{split}
g\mu_B B_{xc} \of{t} 
    &=  \frac{\gamma_L}{2\pi} P_L\ln \abs{\frac{\varepsilon - e \of{V_{DC} + V^{RF}\cos \omega t}} {\varepsilon+U-e\of{V_{DC} +   V^{RF}\cos \omega t}}} \\
    &= \frac{\gamma_L}{2\pi} P_L \ofs{
    \ln \abs{\frac{\varepsilon - eV_{DC}} {\varepsilon+U-eV_{DC}}} + \ln\abs{1+\zeta_0^{-1}\cos\omega t} - \ln\abs{1-\zeta_2^{-1}\cos \omega t}
} 
\end{split}\end{align}
We consider an expresion $\ln\of{1-a\cos \theta}$ for $\abs{a}<1$. 
Defining $r=\frac{a}{1+\sqrt{1-a^2}}$ we have $1-a\cos \theta = \frac{\of{1-re^{i\theta}}\of{1-re^{i\theta}}}{1+r^2} = \frac{1-2r\cos\theta + r^2}{1+r^2}$. 
Citing the identity $\ln\of{1-2r\cos\theta + r^2}=-2\sum_{n=1} \frac{r^n}{n} \cos n\theta$, we arrive at $\ln\of{1-a\cos \theta} = -\ln\of{1+r^2} - 2\sum_{n=1} \frac{r^n}{n} \cos n\theta$. 
With this result, we can write the Fourier expansion of the exchange field in the purely sequential or CB regime $\abs{\zeta_{0,2}} > 1$
\begin{align}\begin{split}
    g\mu_B B_{xc} \of{t} = \frac{\gamma_L}{2\pi} P_L \ofs{
        \ln \abs{\frac{\varepsilon - eV_{DC}} {\varepsilon+U-eV_{DC}}}
        + \ln\frac{1+r_2^2}{1+r_0^2} + 2 \sum_{n=1} \frac{r_2^n - r_0^n} {n} \cos n \omega t 
    } 
\end{split}\end{align}
where $r_{0,2} = \pm \frac{\text{sign}\of{\zeta_{0,2}}}{\abs{\zeta_{0,2}}+ \sqrt{\zeta_{0,2}^2 -1}}$

\section{Computational Details and Parameters}
\label{sec:app-sim-params}

All Calculations were performed using our in-house Python library \href{https://github.com/qphensurf/PyESRSTM.git}{PyESRSTM} \cite{Nachtigall2026PyESRSTM}.
The integrals in Eq.~\eqref{eq:fermi-integral} are calculated using the trapezoidal rule, where the number of points between the integral boundaries $-E_{\rm cut}$ and $E_{\rm cut}$ is selected via a parameter $N_{\rm int}$.

The sum over Bessel functions in Eq.~\eqref{eq:rate-four} is truncated at $m^*$ such that $m^{*}$ is the lowest integer that satisfies ${\rm all}\ k \in \mathcal{Z}, m^* \leq k < m_{\rm max}: Z\abs{\frac{\sum_{m=0}^{k} J_m \of{\frac{eV_{RF}}{\hbar \omega}}}{\sum_{m=0}^{m_{\rm max}} J_m \of{\frac{eV_{\rm RF}}{\hbar\omega}}}} < J_{tol} $ where $J_{\rm tol}$ and $m_{\rm max}$ are convergence parameters.

From Eq.~\eqref{eq:app-remove-bessel} we conclude that frequency $\omega$ only enters the rates as a computation parameter for the sum over the Bessel functions.
We therefore choose a single value $\omega_{\rm rates}$, independent of the driving frequency $\omega$, to evaluate the rates.
Since Eq.~\eqref{eq:app-remove-bessel} is used to derive all analytical expressions in Section~\ref{sec:results}, which agree with numerical calculations almost perfectly, we conclude that this approximation must be valid under conditions considered in this paper. 

All parameters used in this paper, unless specified otherwise, are listed in Table~\ref{tab:app-params}.
We chose a large value for $\tau_C^{-1}$ as it allows for faster convergence with the number of Bessel functions and leads to only a small alteration of numerical results. 
In Section~\ref{sec:results}, we compare the results of numerical calculations with finite $\hbar\tau_C^{-1}$, with an analytical expression where $\tau_C^{-1}$ is zero; the two predictions agree nearly exactly. 
Likewise, since all numerical results agree with their analytical counterparts, it is safe to assume that the numerical calculations are well converged. 

\begin{table}[h]
    \centering
\begin{tabular}{|c|c|c|}\hline
     Parameter& Value & Description  \\\hline
     $\omega_{\rm rates}$ & 20 GHz\* & Frequency used to calculater rates\\ \hline
     $\gamma_R^{0}$ &5 $\mu$eV & Surface coupling strength \\ \hline
     $\gamma_L^{0}$ &1.25 $\mu$eV & Tip coupling strength \\ \hline
     $E_{\rm cut}$ & 300 meV & Cutoff for integrals $I^{\pm}$ \\\hline
     $N_{\rm int}$ & $10^5$ & Number of points for integrals $I^{\pm}$ \\\hline
     $J_{\rm tol}$ & $10^{-4}$ & Convergence parameter of Bessel functions \\\hline
     $\hbar/\tau_C$ & .1 meV\* & QD energy broadening \\\hline
     $\mu^{F}_R$& 0 & Surface Fermi energy\\\hline
     $T$ & 0.5 K& Electrode temperatures \\\hline
     $P_R$& 0 & Surface Polarization\\\hline
     $P_L$& 0.5 & Tip Polarization\\\hline
     $V^{\rm DC}$& 10 meV & DC bias. \\\hline
     $V^{\rm RF}$& 3 meV\* & Oscillating bias on tip\\\hline
     $\varepsilon$ & -10 meV& QD energy \\\hline
     $U$ & 30 meV & Hubbard interaction \\\hline
     $\theta$& 11.3$^\circ$ & Angle between $\vec{B_0}$ and polarization \\\hline
     $B_0$&  0.51 T & Magnetic field strength\\\hline
\end{tabular}
    \caption{Parameters used for the model, unless specified otherwise.}
        \label{tab:app-params}
\end{table}

\end{document}